\documentclass{aa}  

\usepackage{graphicx}
\usepackage{txfonts}
\usepackage{lipsum}
\usepackage{subcaption}         
\usepackage{lscape}             
\usepackage{placeins}           
\usepackage{multirow}
\usepackage{amsmath,amssymb,amsfonts}
\usepackage{mathrsfs}
\usepackage{xcolor}
\usepackage{booktabs}
\usepackage{url}
\usepackage{enumerate}
\usepackage{siunitx}
\usepackage{array}    
\usepackage{dcolumn}  
\usepackage{twemojis}
\usepackage{tikzlings}
                                
\usepackage[colorlinks=true, allcolors=blue]{hyperref}

\newcommand{\gaia}{\emph{Gaia}}
\newcommand{\ges}{\emph{Gaia}-ESO}

\newcommand{\tgex}{\texttt{T-GEX}}

\newcommand{\sn}{\ensuremath{\mathrm{S/N}}}
\newcommand{\feh}{\ensuremath{[\mathrm{Fe}/\mathrm{H}]}}
\newcommand{\mgfe}{\ensuremath{[\mathrm{Mg}/\mathrm{Fe}]}}

\newcommand{\teff}{\ensuremath{T_{\rm eff}}}
\newcommand{\logg}{\ensuremath{\log g}}
\newcommand{\ewha}{\ensuremath{\mathrm{EW}(\mathrm{H}\alpha)}}
\newcommand{\fminha}{\ensuremath{F(\mathrm{H}\alpha)_{\mathrm{min}}}}

\defcitealias{Beltran2026_TGEXII}{Paper II}
\titlerunning{\reflectbox{\twemoji[scale=0.5]{t-rex}} The \tgex\ project}

\begin{document}

   \title{\reflectbox{\twemoji[scale=0.8]{t-rex}} The \tgex\ project}

   \subtitle{I. Stealth UV-bright Sun-like stars in the Galaxy as candidate contributors to the UV upturn}

%
%
%

    \author{
              M. L. L. Dantas\inst{\ref{aff:puc_ia}, \dagger,}\corrauth{mlldantas@protonmail.com; mlldantas@uc.cl}
              \and
              D. Beltrán\inst{\ref{aff:puc_ia}, \dagger}
              \and
              R. E. Giribaldi\inst{\ref{aff:inaf}}
              \and
              R. Smiljanic\inst{\ref{aff:camk}}
              \and
              P. R. T. Coelho\inst{\ref{aff:iag}}
              \and
              P. B. Tissera\inst{\ref{aff:puc_ia}}
              }

   \institute{
              Instituto de Astrofísica, Pontificia Universidad Católica de Chile, Av. Vicuña Mackenna 4860, Santiago, Chile \label{aff:puc_ia}
              \and
              INAF -- Osservatorio Astrofisico di Arcetri, Largo E. Fermi 5, Firenze, 50125, Italy \label{aff:inaf}
              \and
              Nicolaus Copernicus Astronomical Center, Polish Academy of Sciences, ul. Bartycka 18, 00-716, Warsaw, Poland \label{aff:camk}
              \and
              Instituto de Astronomia, Geofísica e Ciências Atmosféricas, Universidade de São Paulo, R. do Matão 1226, São Paulo, 05508-090, Brazil \label{aff:iag}
             }
             
   \date{Received MMMM XX, 20XX}

    \abstract
    {The ultraviolet (UV) excess of old stellar systems is generally attributed to hot evolved stars, interacting binaries, stellar remnants, or residual star formation. Solar-type stars are not expected to emit strongly in the UV, although apparently ordinary main-sequence turn-off (MSTO) stars with unexpected UV emission have been reported.}
    {We investigate the UV properties of Galactic field FGK MSTO stars, their relation to stellar parameters and chromospheric activity, and their possible contribution to quiescent galaxies.}
    {We constructed the Tiny \gaia-ESO + GALEX (\tgex) catalogue by combining \gaia-ESO spectroscopy, \gaia\ DR3 astrometry, GALEX UV photometry, and 2MASS and AllWISE infrared measurements for 37 stars. Astrometric and spectroscopic quality cuts minimise obvious multiplicity and spectral peculiarity, although unresolved companions cannot be excluded. We classified the stars relative to the empirical FGK UV-normal locus, applied a 3-component Gaussian mixture model in the FUV--NUV versus NUV--$G$ plane, measured H$\alpha$ $\lambda 6563$ diagnostics for 24 stars, and performed an IMF-based empirical scaling calculation for three UV-upturn galaxies.}
    {Approximately 2/3 of the sample are UV-abnormal. The GMM identifies a cool, strongly UV-excess group (G1), a hotter UV-normal group (G2), and an intermediate UV-excess group (G3). G1 occupies a narrow, predominantly sub-solar \feh\ range and is $\alpha$-enhanced, with the highest median \mgfe. G3 has the highest median \feh, while G2 spans the broadest metallicity range and has the youngest median age. G1 and G3 extend to old ages, although their youngest estimates are affected by isochrone degeneracy. Enhanced H$\alpha$ core emission occurs only among hotter G2 stars in the available subsample, but only one G1 star has H$\alpha$ coverage. The extragalactic contribution is strongly template-dependent: the median G1 template accounts for at most $\sim20\%$ of the observed UV output, whereas the bluest template could match most or all of the NUV and FUV emission if shared by $\sim10$--$15\%$ of surviving FGK stars.}
    {}


   \maketitle
   \nolinenumbers

   \begingroup
   \renewcommand{\thefootnote}{\ensuremath{\dagger}}
   \footnotetext{Contributed equally to this work.}
   \endgroup

\section{Introduction}
\label{sec:intro}

The Milky Way (MW) is our cosmic Rosetta Stone: the only galaxy in which a large and diverse population of individual stars can be observed and assembled into a coherent narrative of formation and evolution across cosmic time, with precise chrono-chemo-dynamical information and an increasingly comprehensive census of planetary systems \citep[e.g.][]{FBH2002, RixBovy2013, Santos2013, BHG2016, Helmi2020, Dantas2023, Dantas2025a, deLaverny2025, GarciaDelgado2026}. This level of detail transforms questions of galaxy assembly into a resolved reconstruction of physical processes that are otherwise forever blurred into the integrated light of distant galaxies \citep[][]{Walcher2011, Conroy2013}. Yet, one of its most revealing windows—the ultraviolet (UV) regime—remains comparatively underexplored, despite its sensitivity to subtle variations in stellar physics and its ability to expose phenomena invisible at optical and infrared (IR) wavelengths \citep[][]{Martin2005, Bianchi2014, Psaradaki2024}.

Significant UV emission is typically associated with stars hotter than $\sim$ 10\,000 K (such as young OB stars, extreme horizontal-branch and post-asymptotic giant branch stars, interacting binaries, and stellar remnants) while cool stars are expected to contribute negligibly \citep[][]{Brown2000, Rosenfield2012, HPB2014, Eldridge2017, Werle2019}. However, sparse reports challenge this paradigm: a population of UV-bright, cool (effective temperatures \teff $\lesssim 6500$ K) main-sequence (MS) turn-off (MSTO) stars, spectrally akin to the Sun, can exhibit significant UV flux \citep{Smith2014}, echoing earlier attempts to distinguish solar analogues through near-UV discrepancies \citep[see e.g.][]{Hardorp1978, Hardorp1982}.

Low-mass MSTO stars dominate the stellar mass budget of old and quiescent systems, as dictated by the initial mass function \citep[IMF; ][]{Salpeter1955, Kroupa2001, Chabrier2003}, implying that even systematic UV emission from a minority of these stars could alter long-standing interpretations of the UV upturn---the unexpected rise in far-UV (FUV) flux observed in elliptical galaxies and other old stellar systems \citep{CodeWelch1979, Bertola1982, Burstein1988, Dorman1993, Buson2006, Buzzoni2012, Goudfrooij2018, Peacock2018}. What appears anomalous in individual stars may, when integrated across entire galaxies, represent a significant and previously overlooked contributor to this phenomenon.

Here, we combine high-resolution \gaia-ESO spectra \citep{Gilmore2022, Randich2022} with GALEX UV photometry and IR measurements from 2MASS \citep{Skrutskie2006} and AllWISE \citep{Cutri2014} to characterise 37 MSTO stars, of which $\sim$2/3 show anomalous UV emission, forming the Tiny \gaia-ESO + GALEX catalogue (\tgex). These stars present a thermal paradox: by Wien’s displacement law \citep{Wien1893}, photospheres peaking in the optical or near-IR should emit minimally in the UV, particularly at ages $\gtrsim 1$ Gyr \citep{Findeisen2011}. Yet, within the \tgex~sample, we find that the observed UV fluxes exceed the predictions, especially for stars with \teff$\lesssim 5500$ K and median ages $>3$ Gyr. We identify three behaviours: (i) a UV-strong group with FUV--NUV colours comparable to those of UV-upturn galaxies \citep[][]{Oconnell1999, Yi2011}; (ii) a UV-normal population following the expected (FUV--NUV)--\teff\ relation \citep{Smith2014}; and (iii) an intermediate population exhibiting milder but significant UV excesses. The UV-strong and intermediate groups challenge canonical stellar physics and motivate discussion of plausible mechanisms, including unresolved compact companions \citep[][]{Parsons2016}, magnetic or chromospheric activity \citep{Findeisen2011}, He-related evolutionary pathways involving stripped hot components \citep[rather than He-enrichment per se;][]{Oconnell1999}, or opacity-modifying abundance patterns \citep[][]{Iglesias1996, Piau2002, Weiss2006, HBH2021}, and accretion-related processes \citep[][]{Wade1998, Herczeg2008}.

This tripartite classification provides a framework for exploring the physical origins of UV excess in Sun-like dwarfs and highlights how even a modest population of such stars could influence the UV emission of old stellar systems. Linked to the IMF argument above, our results challenge the long-held assumption that UV light in old stellar populations arises solely from hot post-MS phases \citep[][]{Rosenfield2012}, interacting binaries \citep[][]{Han2007, HPB2013, HPB2014, Offner2023}, stellar remnants \citep[][]{Werle2019}, or residual star formation \citep[][]{SalvadorRusinol2020}. The cumulative contribution of cool UV-emitting dwarfs may therefore be non-negligible, particularly in quenched early-type galaxies.

Historically, efforts to understand the UV upturn have focused on unresolved stellar systems, or on dense environments such as globular clusters \citep[][]{Goudfrooij2018, Peacock2018} and anomalous old open clusters \citep[][]{Buson2006, Buzzoni2012}. However, reliance on integrated light or crowded regions limits our ability to test and refine stellar population models at the level required to interpret UV excesses in distant galaxies \citep[][]{Maraston2005, Goncalves2020}. By contrast, MW field stars remain largely unexplored in this context. Here, we take a step toward filling this gap by using resolved Galactic field stars to probe the physics of cool UV emitters and to inform models of UV emission in aged, passively evolving systems.

\section{Data and methodology}
\label{sec:data}

\subsection{The \tgex\ catalogue assembly}
\label{subsec:data_sources}

We assembled a panchromatic stellar catalogue by combining astrometric, photometric, and spectroscopic data from \gaia, \ges, GALEX, 2MASS, and AllWISE. These surveys span a wide range of wavelengths and provide complementary information on stellar positions, motions, and physical parameters.

\gaia\ DR3 \citep{Gaia2023_DR3} delivers high-precision astrometry and photometry for nearly 2 billion stars. We selected sources with positive parallaxes and applied quality cuts following \citet{Fabricius2021}: \texttt{ruwe} $< 1.4$, \texttt{ipd\_frac\_multi\_peak} $\leq 2$, and \texttt{ipd\_gof\_harmonic\_amplitude} $\leq 0.1$. This combination of flags is robust against binarity (i.e. including blue stragglers or blue-stragglers-to-be objects), even close ones \citep{Fabricius2021, Kervella2022}, although we cannot guarantee a sample completely free from those, which will be important and discussed in Sect. \ref{sec:results}.

Spectroscopic data were drawn from the sixth internal data release of the \gaia-ESO survey (iDR6), equivalent to the final public release \citep{Gilmore2022, Randich2022}. Observations were conducted using the Fibre Large Array Multi Element Spectrograph (FLAMES) at the VLT, with targets observed either by UVES ($R \approx 47\,000$) or GIRAFFE ($R \approx 20\,000$). We retained field stars (\texttt{GES\_TYPE = GE\_MW}) with clean spectral flags (\texttt{PECULI = NAN}), removed fast rotators ($v \sin i \geq 10$), and required high signal-to-noise ratios (\sn\ $\geq 40$). Stars missing any of the fundamental parameters (\feh, $\log g$, or \teff) were excluded. Two stars in Group 2 (UV-normal) lack Mg measurements; abundance-based visualisations and correlations involving \mgfe\ therefore use the available measurements only. We refer the reader to \citet{Stonkunte2016} for the selection function of \gaia-ESO.

GALEX UV photometry was included for both NUV and FUV bands, through a tight crossmatch of 2 arcsec with \gaia. The mission provides angular resolutions of approximately 5.3 (NUV) and 4.2 arcsec \citep[FUV;][]{Bianchi2014}. We selected sources with \textsc{SourceExtractor} \citep{BertinArnouts1996} flags \texttt{NUV\_CLASS\_STAR} and \texttt{FUV\_CLASS\_STAR} $\geq 0.7$, ensuring a high likelihood of stellar classification. No \sn\ threshold was imposed to maximise the UV sample, though uncertainties are commensurate to the \sn\; they were used and propagated in subsequent analyses.

Infrared magnitudes from 2MASS \citep[$JHK_s$;][]{Skrutskie2006} and AllWISE \citep[$W1$, $W2$;][]{Cutri2014} were added by crossmatching with \gaia\ positions. We required complete photometric coverage across all five bands for inclusion, as these data are used in age inference (see Sect. \ref{subsec:ages_and_others}).

Our final sample consists of 37 FGK-type stars with UV detections and high-quality spectroscopic and photometric coverage, which we refer to as the \emph{Tiny \gaia-ESO + GALEX} (\tgex) catalogue. Bayesian distances place the \tgex\ stars within the solar neighbourhood, spanning $0.34$--$5.68$ kpc with a median distance of $1.09$ kpc.

\subsection{Stellar ages, extinction, and H$\alpha$ diagnostics}
\label{subsec:ages_and_others}

We estimated stellar ages with \textsc{Unidam} \citep{Mints2017, Mints2018}, which fits PARSEC isochrones \citep{Bressan2012} and requires near- and mid-IR photometry from 2MASS \citep{Skrutskie2006} and AllWISE \citep{Cutri2014}, thereby ensuring panchromatic coverage of our catalogue. Extinction corrections were applied using 3D dust maps via \textsc{Dustmaps} \citep{Green2018_dustmaps, Green2018}. To maximise our sample size, we also queried the value-added catalogues of \citet{Queiroz2023}, who used \textsc{StarHorse} \citep{Queiroz2018} to infer stellar parameters (including ages) for several spectroscopic surveys, including \gaia-ESO. Only one additional star in our sample had an age reported in this compilation, which we retained.

To extract homogeneous H$\alpha$ diagnostics beyond the \gaia-ESO pipeline, we re-analysed the 24 spectra in the \tgex\ sample with wavelength coverage over 5822.00--6830.91~\AA, which includes H$\alpha$. The remaining 13 stars lack suitable H$\alpha$ coverage. Importantly, the available subsample contains only one Group~1 star, substantially limiting our ability to assess chromospheric activity among the most UV-extreme objects. We continuum-normalised each available spectrum with \textsc{Rassine} \citep{Cretignier2020}, an open-source \textsc{Python} tool that reconstructs the stellar continuum assuming a 1D atmosphere, using convex-hull and alpha-shape algorithms. The code was run in non-interactive mode with default settings (including \texttt{par\_smoothing\_box = 6}, \texttt{par\_stretching = auto\_0.5}, and \texttt{par\_fwhm = auto}) and applied uniformly through a custom wrapper that iteratively updated the configuration file and executed the code for each spectrum. From the normalised profiles, we measured the equivalent width of the H$\alpha$ absorption line [\ewha] and its minimum flux [\fminha].

These H$\alpha$ measurements provide a deliberately coarse and incomplete proxy for chromospheric activity and line morphology. We use them to examine whether H$\alpha$ behaviour broadly tracks GALEX UV magnitudes and colours within the 24-star subsample. Because the H$\alpha$ coverage is uneven across the GMM groups, particularly for Group~1, the resulting trends should be regarded as descriptive rather than as a definitive test of the origin of the UV excess.

\subsection{Sample clustering}
\label{subsec:clustering}

We applied a Gaussian Mixture Model \citep[GMM;][but see also \citealt{deSouza2017} for an example of GMM application in astronomy, as well as the description of the workings of a GMM]{McLachlan2000, Hastie2001, Murphy2013} to partition the stellar UV--optical colour space. A GMM represents the data as a superposition of multivariate Gaussian components and provides a flexible, probabilistic way to assign stars to colour-space regimes. Here, however, the GMM is used as a phenomenological classification tool rather than as a formal model-selection exercise.

The model was applied to the FUV--NUV versus NUV--$G$ colour--colour diagram, which we adopt as a stellar analogue of the classical FUV--NUV versus NUV--$r$ space used by \citet{Yi2011} to separate residual star-forming, UV-weak, and UV-upturn galaxies. We used \gaia's $G$ band ($\lambda_{\rm eff} \sim$ 6730~\AA) as an approximate replacement for the SDSS $r$-band \citep[$\lambda_{\rm eff} \sim$ 6220~\AA;][]{York2000}.

Because the \tgex\ stars occupy an approximately elongated sequence in this plane, information criteria such as the Bayesian Information Criterion \citep[BIC;][]{Schwarz1978} favour the simplest one-component description when the number of components is allowed to vary freely. We nevertheless adopted three components to provide an approximate, data-driven counterpart to the three UV regimes commonly used in unresolved-galaxy studies, while avoiding the imposition of fixed straight-line boundaries on the stellar sample. We adopted full covariance matrices, allowing each component to have its own orientation and elongation in the colour--colour plane.

Within this phenomenological framework, the adopted three-component GMM cleanly separates the sample into three visually and physically interpretable UV-colour regimes: sources with UV colours comparable to classical UV-upturn systems, UV-normal stars as defined by \citet[][see also Fig. \ref{fig:fuv_nuv_teff_kiel}, left panel]{Smith2014}, and a transitional UV-excess regime with intermediate colours. These groupings are examined in more detail in Sect. \ref{sec:stellar_properties}.

\subsection{A feasibility test for FGK-star contributions to the UV upturn}
\label{subsec:method_brute_sed_fit}

We developed an empirical scaling calculation to estimate the fraction of a quiescent galaxy's UV luminosity that could be matched by UV-bright FGK stars. The analysis is anchored to the UV-upturn galaxy sample presented in \citet{Dantas2020, Dantas2021}, drawn from the GAMA DR3 survey \citep{Driver2009, Baldry2018} with reliable GALEX GR6/7 \citep{Martin2005} and SDSS DR7 photometry \citep{York2000}. Galaxy classification follows the UV--optical colour criteria of \citep{Yi2011}, which require both UV and optical bands to distinguish (residual) star-forming, UV-weak, and UV-upturn systems. To span the observed range of UV-upturn strengths, we rank the UV-upturn population by a UV-brightness score derived from GALEX and SDSS magnitudes, divide the sample into terciles, and randomly select one quiescent galaxy from each tercile using a fixed random seed. These three systems are intended as representative case studies rather than a statistical census of the full population.

For each selected galaxy, we perform an empirical UV-luminosity scaling calculation rather than a full SED fit. The calculation asks how much of the observed galaxy FUV and NUV luminosity could be matched by a hypothetical UV-excess FGK component, if a fraction of the surviving FGK population had UV luminosities comparable to those measured in \tgex. We therefore do not model or subtract the UV contributions from other stellar evolutionary channels. Instead, these contributions remain included in the observed galaxy UV luminosity, which serves as the denominator of the calculation. The numerator is obtained by combining empirically measured per-star UV luminosities from \tgex\ with a Kroupa initial mass function \citep{Kroupa2001}, a formed stellar mass, and an age-dependent estimate of the number of surviving FGK stars. The resulting quantity is a physically transparent feasibility estimate: it measures the fraction of the observed UV output that would be matched by UV-bright FGK-like systems under the adopted assumptions.

We adopt the Kroupa IMF, defined as a broken power law,

\begin{align}
    \xi(M) \equiv \frac{{\rm d}N}{{\rm d}M} \propto 
    \begin{cases}
        M^{-\alpha_1}, & 0.08 \le M/M_\odot < 0.50,\\
        M^{-\alpha_2}, & 0.50 \le M/M_\odot \le 100,
    \end{cases} \notag \\ 
    (\alpha_1, \alpha_2) = (1.3,\,2.3),
    \label{eq:kroupa_imf}
\end{align}

\noindent where $\xi(M) = {\rm d}N/{\rm d}M$ is the differential stellar number function, such that $\xi(M)\,{\rm d}M$ gives the number of stars in the mass interval $(M,\,M+{\rm d}M)$. Continuity is enforced at the mass breaks to ensure a smooth IMF. A single normalisation constant (defined in Eq. \ref{eq:imf_norm}) scales the function to the total formed stellar mass of the population.

Throughout, we assume a universal Kroupa-like IMF consistent with the MW field population. Alternative parameterisations such as those by \citet{Chabrier2003} or \citet{Salpeter1955} yield nearly identical relative weights at the low-mass end relevant for FGK stars and therefore have a negligible impact on our results.

\subsubsection{IMF normalisation and formed mass}
\label{subsubsec:imf_norm}

The observed stellar mass of a galaxy corresponds to the fraction of the initially formed stellar mass that remains locked in stars and stellar remnants at the present epoch. We express this relation as

\begin{equation}
    M_{\star}^{\rm present} = M_{\star}^{\rm formed}\,(1 - R),
    \label{eq:formed_mass}
\end{equation}

\noindent where $M_{\star}^{\rm present}$ is the present-day stellar mass taken from the GAMA catalogue for the selected galaxies, $M_{\star}^{\rm formed}$ is the inferred total mass converted into stars at birth, and $R$ is the returned-mass fraction, i.e. the proportion of stellar material recycled into the interstellar medium through winds and supernovae. Thus, the present-day stellar masses are not assumed in our calculation; the assumption enters when converting these catalogue masses into initially formed stellar masses through the adopted value of $R$.

For a Kroupa-like IMF and stellar populations older than a few Gyr, stellar-evolution and population-synthesis models predict returned fractions in the range $R \sim 0.2$--$0.4$ \citep[][]{Leitner2011, Lehnert2015, Vincenzo2016}. We therefore adopt a representative value of $R = 0.4$ \citep[][]{Vincenzo2016}. Varying $R$ across the full range $0.2 \le R \le 0.4$ changes the inferred formed stellar mass, and hence the absolute normalisation of the surviving FGK population, by $\sim$30--40\%; this rescaling does not significantly affect the qualitative trends explored here, which are dominated by the assumed UV-excess FGK fraction and the intrinsic scatter in stellar UV luminosities. In practice, this effect is smaller than the galaxy-to-galaxy UV luminosity differences within the UV-upturn sample itself. Uncertainty in $R$ is propagated through the bootstrap resampling described in Sect. \ref{subsubsec:bootstrap_method}, where $R$ is drawn from a normal distribution and restricted to physically plausible values.

The corresponding normalised IMF is written as

\begin{equation}
    \xi_{\rm n}(M) = A\,\xi(M), 
    \qquad
    M_{\star}^{\rm formed} = \int_{M_{\rm min}}^{M_{\rm max}} M\,\xi_{\rm n}(M)\,{\rm d}M,
    \label{eq:imf_norm}
\end{equation}

\noindent where $\xi_{\rm n}(M)$ gives the number of stars per unit stellar mass interval, and $A$ is the normalisation constant that ensures mass conservation over the adopted limits $M_{\rm min}=0.08\,M_\odot$ and $M_{\rm max}=100\,M_\odot$. In practice, $A$ scales the unnormalised IMF $\xi(M)$ (Eq. \ref{eq:kroupa_imf}) such that its mass-weighted integral equals $M_{\star}^{\rm formed}$. This formulation allows all subsequent quantities, including the surviving FGK population, to be expressed directly in terms of the observable $M_{\star}^{\rm present}$.

\subsubsection{Surviving stars at age $\tau$}
\label{subsubsec:surviving_stars}

We approximate each selected quiescent galaxy by a representative old stellar population age $\tau$, used only to estimate which FGK stars remain on the MS. This should not be interpreted as a full reconstruction of the galaxy star-formation history. For this representative age, the MS turn-off mass $M_{\rm to}(\tau)$ is defined by $t_{\rm MS}\!\left(M_{\rm to}\right)=\tau$, i.e. stars with $M>M_{\rm to}$ have exhausted core hydrogen and no longer contribute to the MS FGK census. We adopt the standard scaling that the MS lifetime is of order the nuclear timescale, $t_{\rm MS}\propto M/L$ (fuel divided by luminosity; see  \citet{Hansen2004}, Sect.~1.7 therein; see also \citet{Hurley2000} and the Population~I lifetime fit of \citet{Bahcall1983}), and approximate it as a power law in mass,

\begin{equation}
    t_{\rm MS}(M) \simeq 10~{\rm Gyr}\,\left(\frac{M}{M_\odot}\right)^{-\eta},
    \quad \Rightarrow \quad
    M_{\rm to}(\tau) \simeq M_\odot \left(\frac{\tau}{10~{\rm Gyr}}\right)^{-1/\eta},
    \label{eq:turnoff}
\end{equation}

\noindent where the exponent $\eta$ encapsulates the (mass-dependent) MS mass--luminosity relation. Textbook estimates give $\eta \approx 2.5$--3.5 across the approximate solar-mass regime, with steeper scalings applying toward the upper MS. We adopt $\eta=2.5$ as a fiducial value and treat $\eta$ as a nuisance parameter by sampling it in the bootstrap over the plausible range for solar-type stars (Sect.~\ref{subsubsec:bootstrap_method}), thereby propagating uncertainties in the mass--lifetime relation.

The number of surviving FGK stars at age $\tau$ is then

\begin{equation}
    N_{\rm FGK}^{\rm surv}(\tau) = \int_{M_{\rm lo}}^{M_{\rm hi}(\tau)} \xi_{\rm n}(M)\,{\rm d}M, \\ 
    M_{\rm hi}(\tau) \equiv \min\!\left[M_{\rm to}(\tau),\,M_{\rm FGK,hi}\right],
    \label{eq:n_fgk_surv}
\end{equation}

\noindent where $M_{\rm lo}\equiv \max(M_{\rm FGK,lo}, M_{\rm min})$. We adopt $M_{\rm FGK,lo}=0.5 M_\odot$ and $M_{\rm FGK,hi}=1.1 M_\odot$ as a restricted mass window for the cool FGK/MSTO population represented by the strongest UV-excess \tgex\ templates, rather than as an exhaustive mapping of all possible F--K spectral types. This choice is motivated by the fact that the most UV-excess \tgex\ stars are cool and their mass is lower; however, given the small size and selection-function-limited nature of the sample, we cannot determine whether comparable UV-excess behaviour also occurs among more massive early-F stars. Equation \eqref{eq:n_fgk_surv} ensures that the most massive stars within this adopted window are automatically removed at older ages when $M_{\rm to}(\tau) < M_{\rm FGK,hi}$. A broader FGK mass interval would change the number of potential contributors, but its effect on the inferred UV-excess fraction would depend on whether those additional stars exhibit comparable UV-excess behaviour. We therefore treat the adopted mass window as a simplifying assumption of the feasibility test rather than as a complete census of all UV-bright FGK contributors. The numerical scale of the three representative UV-upturn galaxies, together with the resulting surviving FGK counts for the fiducial calculation, is summarised in Table \ref{tab:selected_uvupturn_galaxies} in the Appendix.

For the fiducial calculation, the selected galaxies contain $N_{\rm FGK}^{\rm surv}\simeq(1.1$--$3.3)\times10^{10}$ surviving stars in the adopted FGK mass window. Because the scaling calculation uses catalogue absolute AB magnitudes, galaxy distances enter only through the upstream construction of $M_{\rm FUV}$ and $M_{\rm NUV}$ and are not used as an additional variable in Eq.~\eqref{eq:fgk_fraction}.

\subsubsection{Per-star UV templates and the UV-excess FGK component}
\label{subsubsec:per_star_uv_fgk_lum}

Let $L_{\rm FUV}^{\star}$ and $L_{\rm NUV}^{\star}$ denote the per-star UV luminosities assigned to an empirical \tgex\ UV-excess template. In practice, we evaluate three templates derived from \tgex\ Group~1: the individual stars with the reddest and bluest FUV--NUV colours, and a median template constructed from the Group~1 median FUV and NUV absolute magnitudes. These templates are used only through their assigned UV luminosities; their individual stellar ages are not used in the extragalactic scaling. This distinction is particularly important for the bluest template, which has a formal isochrone age of $\sim$200 Myr but lies in an age-degenerate region of the Kiel diagram.

Let $f_{\rm \textsc{uvx}}$ denote the fraction of surviving FGK stars assumed to belong to this UV-excess population. The UV luminosity associated with this component in a population of age $\tau$ is then

\begin{align}
    L_{\rm FUV}^{\rm FGK}(\tau, f_{\rm \textsc{uvx}}) &=
    f_{\rm \textsc{uvx}}\,N_{\rm FGK}^{\rm surv}(\tau)\,L_{\rm FUV}^{\star}, \notag \\
    L_{\rm NUV}^{\rm FGK}(\tau, f_{\rm \textsc{uvx}}) &=
    f_{\rm \textsc{uvx}}\,N_{\rm FGK}^{\rm surv}(\tau)\,L_{\rm NUV}^{\star},
    \label{eq:fgk_uv_lum}
\end{align}

\noindent where $N_{\rm FGK}^{\rm surv}(\tau)$ is given by Eq.~\eqref{eq:n_fgk_surv}. By construction, this formulation treats the \tgex\ stars as empirical UV templates for the UV-excess subset of the FGK population, rather than as theoretical stellar-atmosphere predictions.

We estimate $L_{\rm FUV}^{\star}$ and $L_{\rm NUV}^{\star}$ by converting GALEX AB magnitudes to monochromatic luminosities \citep[see ][]{Bianchi2014} according to

\begin{align}
    L_\nu = 4\pi\,d^2\,f_\nu, \\
    m_{\rm AB} = -2.5\log_{10}\!\left(\frac{f_\nu}{3631~{\rm Jy}}\right), \\
    M_{\rm AB} = m_{\rm AB} - 5\log_{10}\!\left(\frac{d}{10~{\rm pc}}\right),
    \label{eq:ab_system}
\end{align}

\noindent such that $L_\nu \propto 10^{-0.4\,M_{\rm AB}}$. For galaxy-integrated quantities, we use absolute AB magnitudes and convert them to $L_\nu$ using the appropriate band zero-points. All stellar (Sect. \ref{subsec:ages_and_others}) and galaxy magnitudes \citep[see][]{Dantas2020, Dantas2021} have been corrected for foreground extinction. The empirical template magnitudes and corresponding monochromatic luminosities used in Eq. \eqref{eq:fgk_uv_lum} are listed in Table \ref{tab:tgex_uv_templates} available in the Appendix.

\subsubsection{Fraction of the observed UV emission}
\label{subsubsec:fgk_uv_fraction}

For each galaxy, we convert the observed absolute magnitudes to UV luminosities, $L_{\rm FUV}^{\rm obs}$ and $L_{\rm NUV}^{\rm obs}$. The fractional contribution of the UV-excess FGK component to the observed UV emission is then defined as

\begin{equation}
    f_{\rm \textsc{fgk,uv}}(\tau, f_{\rm \textsc{uvx}}) \equiv 
    \frac{L_{\rm UV}^{\rm FGK}(\tau, f_{\rm \textsc{uvx}})}
         {L_{\rm UV}^{\rm obs}},
    \label{eq:fgk_fraction}
\end{equation}

\noindent computed separately for the FUV and NUV bands. This quantity should not be interpreted as a fitted decomposition of the galaxy SED. It is the fraction of the already-observed UV luminosity that would be matched by the assumed UV-excess FGK component for a given IMF, population age, returned-mass fraction, empirical UV template, and value of $f_{\rm \textsc{uvx}}$.

Values of $f_{\rm FGK,UV}$ approaching unity indicate that the assumed UV-excess FGK component can match most of the observed UV luminosity under the adopted assumptions, whereas smaller values imply a sub-dominant contribution in that band.

\subsubsection{Framework for estimating the FGK-driven UV emission}
\label{subsubsec:putting_it_together}

For each galaxy, we estimate the total formed stellar mass $M_{\star}^{\rm formed}$ from the present-day stellar mass $M_{\star}^{\rm present}$ using Eq.~\eqref{eq:formed_mass}. The IMF is then normalised according to Eq.~\eqref{eq:imf_norm}, and the number of surviving FGK stars at population age $\tau$ is computed via Eqs.~\eqref{eq:turnoff} and \eqref{eq:n_fgk_surv}. We then assume that a fraction $f_{\rm \textsc{uvx}}$ of these surviving FGK stars belongs to the UV-excess component represented by the empirical \tgex\ templates. The resulting UV luminosity is obtained using Eq.~\eqref{eq:fgk_uv_lum}.

Uncertainties are propagated through bootstrap realisations for the selected galaxies, sampling the returned-mass fraction $R$, the MS lifetime exponent $\eta$, the galaxy ages, and the per-star UV luminosity distribution. When stellar ages are unavailable, $\tau$ is sampled uniformly over $5$--$13$ Gyr. For each realisation, we compute the fraction of the observed UV emission matched by the UV-excess FGK component, $f_{\rm FGK,UV}$ [Eq. \eqref{eq:fgk_fraction}], and summarise the resulting distributions using their median and 16th--84th percentile ranges.

\subsubsection{Bootstrap resampling and uncertainty propagation}
\label{subsubsec:bootstrap_method}

Uncertainties in the FGK-driven UV contribution are quantified through bootstrap realisations that propagate uncertainties in the quantities entering the scaling calculation: the returned-mass fraction $R$, the MS lifetime exponent $\eta$, the adopted galaxy ages from the GAMA value-added catalogues, and the empirical UV luminosities of the \tgex\ templates. Each bootstrap realisation produces a self-consistent instance of the scaling calculation, from which the fraction of the observed UV emission matched by the UV-excess FGK component, $f_{\rm FGK,UV}$, is recomputed following Eq.~\eqref{eq:fgk_fraction}.

In each iteration, the returned-mass fraction $R$ entering Eq.~\eqref{eq:formed_mass} is drawn from a normal distribution, $R \sim \mathcal{N}(0.4,\,0.05)$, consistent with expectations for old stellar populations under Kroupa-like initial mass functions \citep[][]{Vincenzo2016}. The MS lifetime exponent $\eta$ in Eq.~\eqref{eq:turnoff} is sampled from a normal distribution, $\eta \sim \mathcal{N}(2.5,\,0.3)$, reflecting uncertainties in the mass--lifetime relation across the FGK mass regime and truncated to physically plausible values for solar-type stars.

Diversity among the empirical UV templates is incorporated by resampling, with replacement, the individual \tgex\ Group~1 stars and their GALEX FUV and NUV absolute magnitudes. These resampled stellar luminosities are used only to compute the UV luminosity of the assumed UV-excess FGK component in Eq.~\eqref{eq:fgk_uv_lum}; they are not inserted into or fitted to a galaxy spectrum or SED. The UV light from all non-FGK sources remains part of the observed galaxy luminosity $L_{\rm UV}^{\rm obs}$ in the denominator of Eq.~\eqref{eq:fgk_fraction}. This procedure captures both the intrinsic scatter in per-star UV luminosities and the sensitivity of the result to the extreme UV-bright tail of the FGK population.

For each galaxy, the stellar population age $\tau$ is perturbed within its uncertainty by drawing uniformly within $\pm1$~Gyr of the adopted age estimate when available. When no reliable age constraint exists, $\tau$ is drawn uniformly over the range $5$--$13$~Gyr, representative of old, quiescent stellar populations.

We perform 2\,000 bootstrap realisations. From the resulting distributions of $f_{\rm FGK,UV}$, we compute the median as the central estimate, the 16th--84th percentile interval as an approximate $1\sigma$ confidence range, the 2.5th--97.5th percentile interval as an approximate $2\sigma$ range, and the median absolute deviation (MAD) as a robust measure of dispersion. These statistics define the central values and uncertainty envelopes shown in Fig.~\ref{fig:nuv_fuv_contributions}.

\section{Results}
\label{sec:results}

\subsection{Stellar properties and UV behaviour in the \tgex\ sample}
\label{sec:stellar_properties}

\begin{figure*}
    \centering
    \includegraphics[width=\linewidth,trim={3mm 7mm 3mm 7mm},clip]{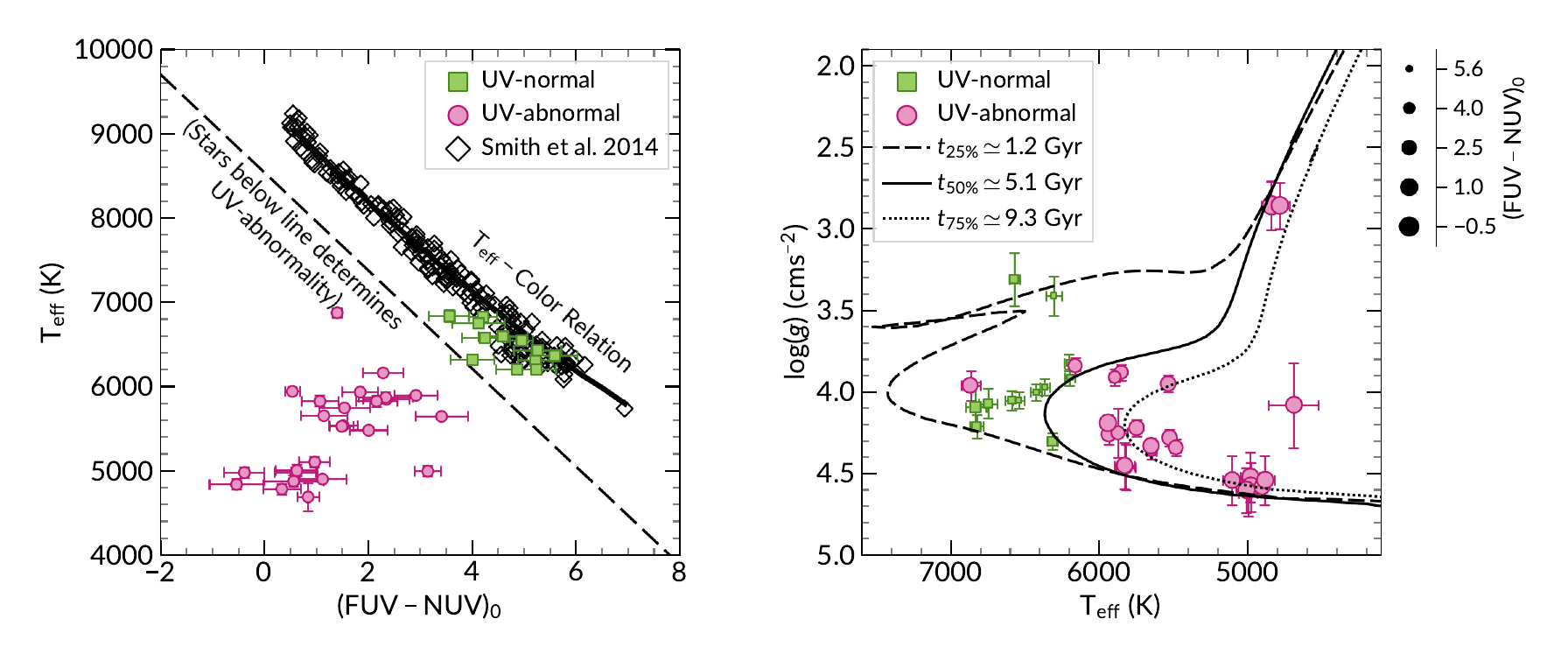}
    \caption{UV-colour versus \teff\ relation and Kiel diagram. Left panel: \teff\ versus FUV–NUV colour for our stellar sample. Black diamonds denote the UV-normal stars from \citet{Smith2014}, together with their defined UV-normality trend (solid line) and threshold (dotted line). The numerical data from \citet{Smith2014} were retrieved from the published figures using \textsc{WebPlotDigitizer} \citep{RohatgiWebPlotDigitizer}. Our sample is classified accordingly, with UV-normal stars in green and UV-abnormal stars in pink. Right panel: Kiel diagram (\teff\ versus surface gravity, \logg) of UV-normal (green) and UV-abnormal (pink) stars, with marker sizes scaling inversely with FUV-NUV (bigger markers indicate larger UV excess). Evolutionary tracks from the PARSEC isochrones \citep{Bressan2012} are shown (at fixed median stellar \feh) show median stellar age (solid line), with 25th and 75th percentiles as dotted and dashed lines, respectively.}
    \label{fig:fuv_nuv_teff_kiel}
\end{figure*}

Figure~\ref{fig:fuv_nuv_teff_kiel} summarises the UV behaviour of stars in the \tgex\ sample. In the FUV--NUV versus \teff\ plane (left panel), we apply the UV-normality criterion of \citet{Smith2014}. Approximately $2/3$ of the \tgex\ stars fall below this threshold and are classified as UV-abnormal, while the remaining third are UV-normal. The UV-abnormal population includes a substantial number of cool stars (\teff\ $\lesssim$ 5000 K) that exhibit increasing FUV excess toward lower \teff.

The Kiel diagram (right panel) shows that both UV-normal and UV-abnormal stars are predominantly MS objects. UV-normal stars tend to occupy slightly higher \teff, while UV-abnormal stars generally display stronger FUV dominance, as reflected by larger marker sizes. Age estimates derived from isochrone fitting are discussed in Sect. \ref{subsec:ages_and_others}.

\begin{figure*}
    \centering
    \includegraphics[width=\linewidth,trim={3mm 7mm 3mm 7mm},clip]{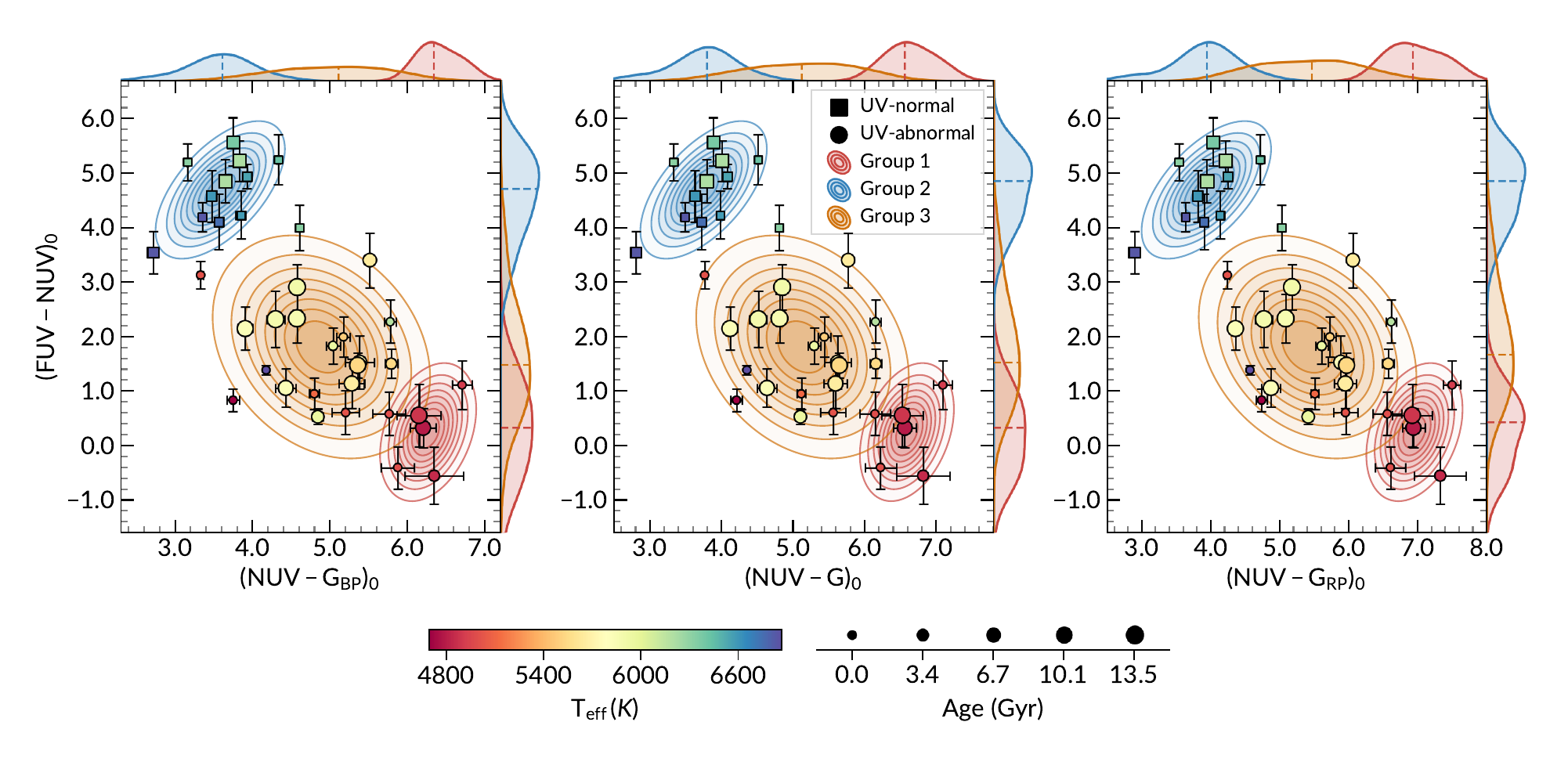}
    \caption{Panels (left to right): Colour--colour diagrams combining GALEX (FUV, NUV) and \gaia\ ($G_{\rm BP}$, $G$, $G_{\rm RP}$) photometry. The Gaussian mixture model (GMM) is fitted in the central (FUV--NUV versus NUV--$G$) plane, and the resulting group contours are projected consistently onto the adjacent colour combinations. Marker shape indicates UV classification (squares: UV-normal; circles: UV-abnormal), marker size scales with stellar age, and colour encodes \teff. Marginal kernel density distributions are shown for each colour axis. The GMM identifies three distinct stellar groups within the \tgex\ sample, whose numbering follows the GMM output and is retained throughout the analysis. The colour scale encodes \teff, with the coolest stars concentrated mainly in Group~1, the hottest stars in Group~2, and intermediate-temperature UV-abnormal stars in Group~3; the corresponding \teff\ distributions are shown explicitly in Fig.~\ref{fig:params_distribution}.}
    \label{fig:color_color_gmm}
\end{figure*}

Figure \ref{fig:color_color_gmm} shows the results of GMM applied in the FUV--NUV versus NUV--$G$ colour space (central panel). The group boundaries derived in this space are projected consistently onto adjacent colour--colour diagrams using alternative \gaia\ bands ($G_{\rm BP}$ and $G_{\rm RP}$). The GMM identifies three distinct groups within the \tgex\ sample. Group 2 contains nearly all UV-normal stars, with only one UV-normal outlier assigned to Group 3. Conversely, Groups 1 and 3 are populated exclusively by UV-abnormal stars. This demonstrates that the UV classification derived in Fig. \ref{fig:fuv_nuv_teff_kiel} is preserved in a higher-dimensional UV--optical colour space. A clear \teff\ gradient is evident across the three groups. Group 2 corresponds to the hottest stars in the sample, Group 1 contains the coolest stars despite occupying the bluest FUV-NUV, and Group 3 spans intermediate \teff\ while remaining UV-abnormal.

\begin{figure*}
    \centering
    \includegraphics[width=\linewidth,trim={3mm 7mm 3mm 7mm},clip]{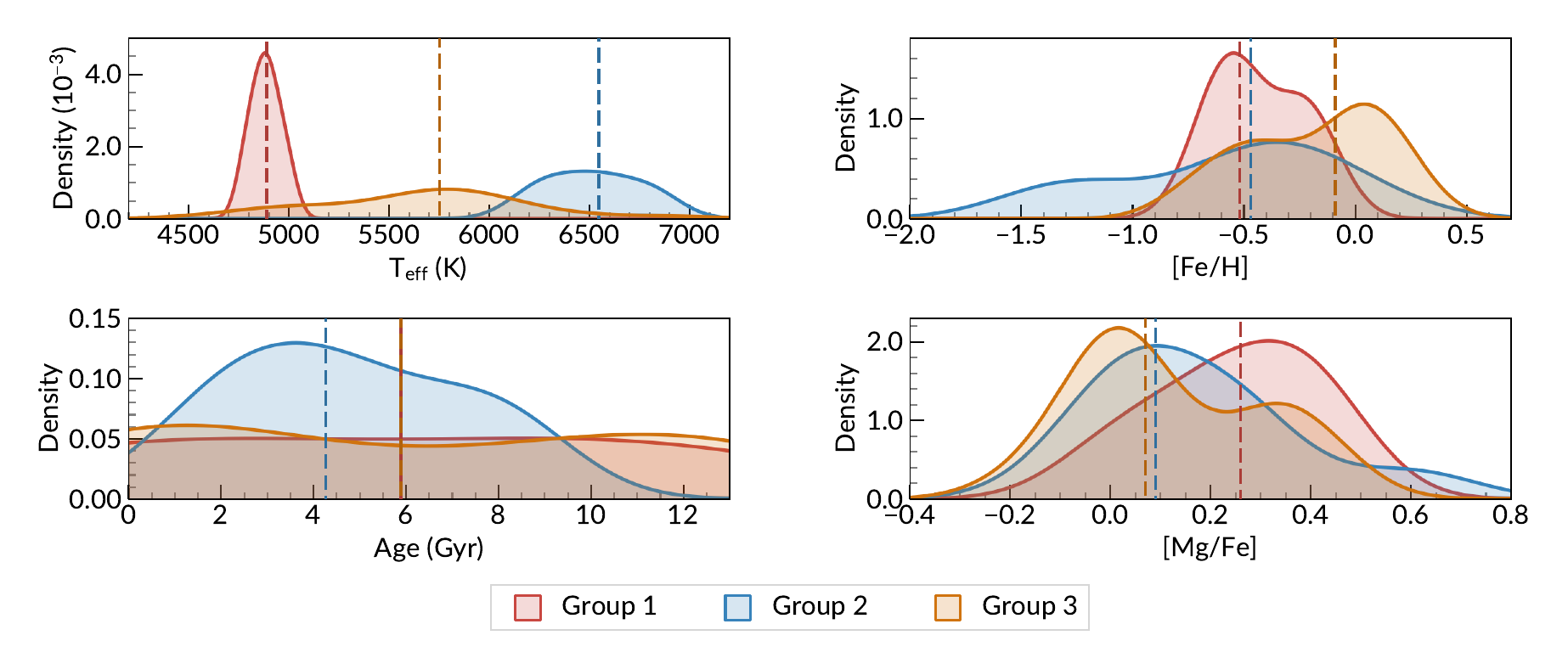}
    \caption{Panels (top-left to bottom-right): Gaussian kernel density distributions of key stellar parameters: \teff, \feh, age, and \mgfe. Vertical dashed lines mark the median (50th percentile) of each distribution, shown separately for each GMM group defined in Fig. \ref{fig:color_color_gmm}. The detailed descriptive statistics of these distributions are displayed in Table \ref{tab:gmms_params} in the Extended Data section. The \mgfe\ panel excludes two UV-normal Group 2 stars without available Mg abundances. These distributions are intended as a descriptive view of the GMM-defined groups within the \tgex\ sample and should not be interpreted as intrinsic population distributions independent of sample selection.}
    \label{fig:params_distribution}
\end{figure*}

Figure \ref{fig:params_distribution} shows the distributions of \teff, \feh, age, and \mgfe\ for the three GMM groups. The \teff\ distributions are sharply separated, despite \teff\ not being used as an input to the GMM, showing that the UV-colour partition is accompanied by a strong temperature gradient.

Group 2 spans the broadest range in \feh, from very metal-poor to super-metal-rich values, whereas the Groups 1 and 3 occupy narrower metallicity ranges within the present sample. Group 3 has the highest median \feh, while Groups 1 and 2 have similar medians despite different metallicity ranges. In terms of \mgfe, Group 1 appears $\alpha$-enhanced relative to the other groups, while Group 3 shows a secondary higher-\mgfe\ peak. These descriptive trends suggest that the UV-abnormal groups do not simply reproduce the same chemical distribution as the UV-normal stars.

Stellar ages span a wide range in all groups. Groups 1 and 3 have a mixture of young and old GK-type stars with similar median ages of $\sim$7 Gyr, while Group 2 shows a younger median age of $\sim$4 Gyr, although age uncertainties are larger due to isochrone degeneracy in this region of the Kiel diagram (as shown in Fig. \ref{fig:fuv_nuv_teff_kiel}, right panel, and Table \ref{tab:gmms_params}).

\subsection{H$\alpha$ diagnostics of chromospheric activity}
\label{subsec:tgex_chromo}

\begin{figure}
    \centering
    \includegraphics[width=\linewidth,trim={3mm 7mm 3mm 7mm},clip]{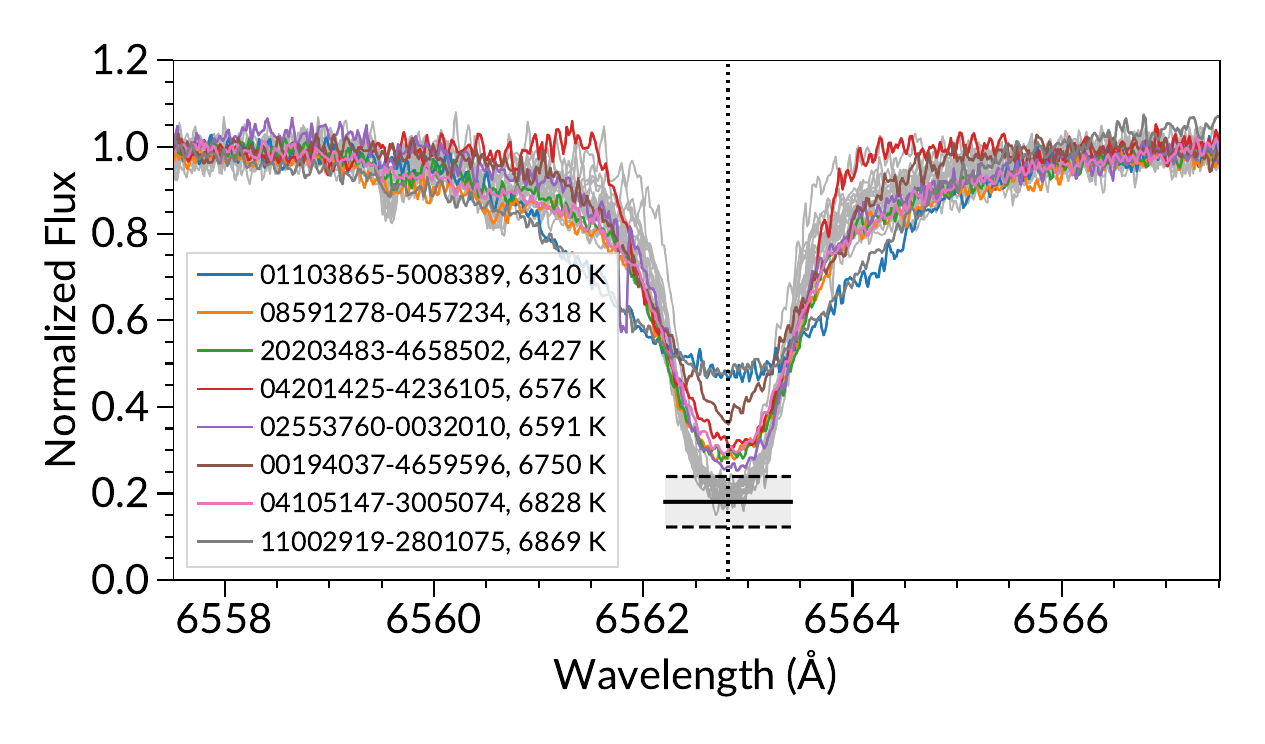}
    \caption{H$\alpha$ line cores for the 24 \tgex\ stars with suitable spectral coverage. The line centre is marked by the vertical dashed line. The most frequent core flux level ($f_{\rm core}=0.18$) is shown by the horizontal solid black line, with the corresponding $\pm3\sigma$ interval ($\sigma=0.02$) indicated by dashed black lines. Stars whose H$\alpha$ core flux exceeds this range are highlighted with coloured profiles. Within the available subsample, these stars belong to Group~2 and are predominantly hot (\teff\ $\gtrsim 6000$ K). Each highlighted spectrum is annotated with its \gaia-ESO identifier (\texttt{CNAME}) and \teff. Only one Group~1 star has H$\alpha$ coverage, preventing a representative comparison with the UV-extreme population. Spectra are shown shifted to the H$\alpha$ rest wavelength (6562.8~\AA).}
    \label{fig:Halpha}
\end{figure}

Chromospheric activity has long been recognised as a potential contributor to the UV emission of FGKM stars, motivating its examination in the context of the \tgex\ UV-abnormal population \citep{Smith2010, Findeisen2011}. We use H$\alpha$, the strongest Balmer line accessible in the relevant \gaia-ESO spectra, as an activity diagnostic. Suitable H$\alpha$ coverage is available for 24 of the 37 \tgex\ stars. This coverage is uneven across the GMM groups and includes only one Group~1 star.

Figure \ref{fig:Halpha} shows that, within the available subsample, stars with elevated H$\alpha$ core flux belong to Group~2. These stars are predominantly hot and have young-to-intermediate ages (7 out of 12 having ages $<5$~Gyr), consistent with the expectation that chromospheric activity is more readily detectable at higher \teff\ and younger ages \citep[][]{Souza2024MNRAS.532..563S}. H$\alpha$ variations in older and cooler stars may also be subtler and more difficult to detect at the \sn\ level of the present data.

None of the observed UV-abnormal stars shows comparably enhanced H$\alpha$ core emission. This suggests that strong H$\alpha$ activity is not ubiquitous among the UV-abnormal stars with available measurements. However, the limited and uneven coverage, particularly the presence of only one observed Group~1 star, prevents us from determining whether chromospheric activity contributes to the UV excess of the broader population. The \gaia-ESO wavelength coverage also does not include the \ion{Ca}{II} H and K lines, precluding a complementary and potentially more sensitive activity diagnostic.

\subsection{Global correlations among stellar parameters in UV-bright FGK stars}
\label{subsec:chord_global_corr}

\begin{figure}
    \centering
    \includegraphics[width=\linewidth,trim={5mm 17mm 5mm 7mm},clip]{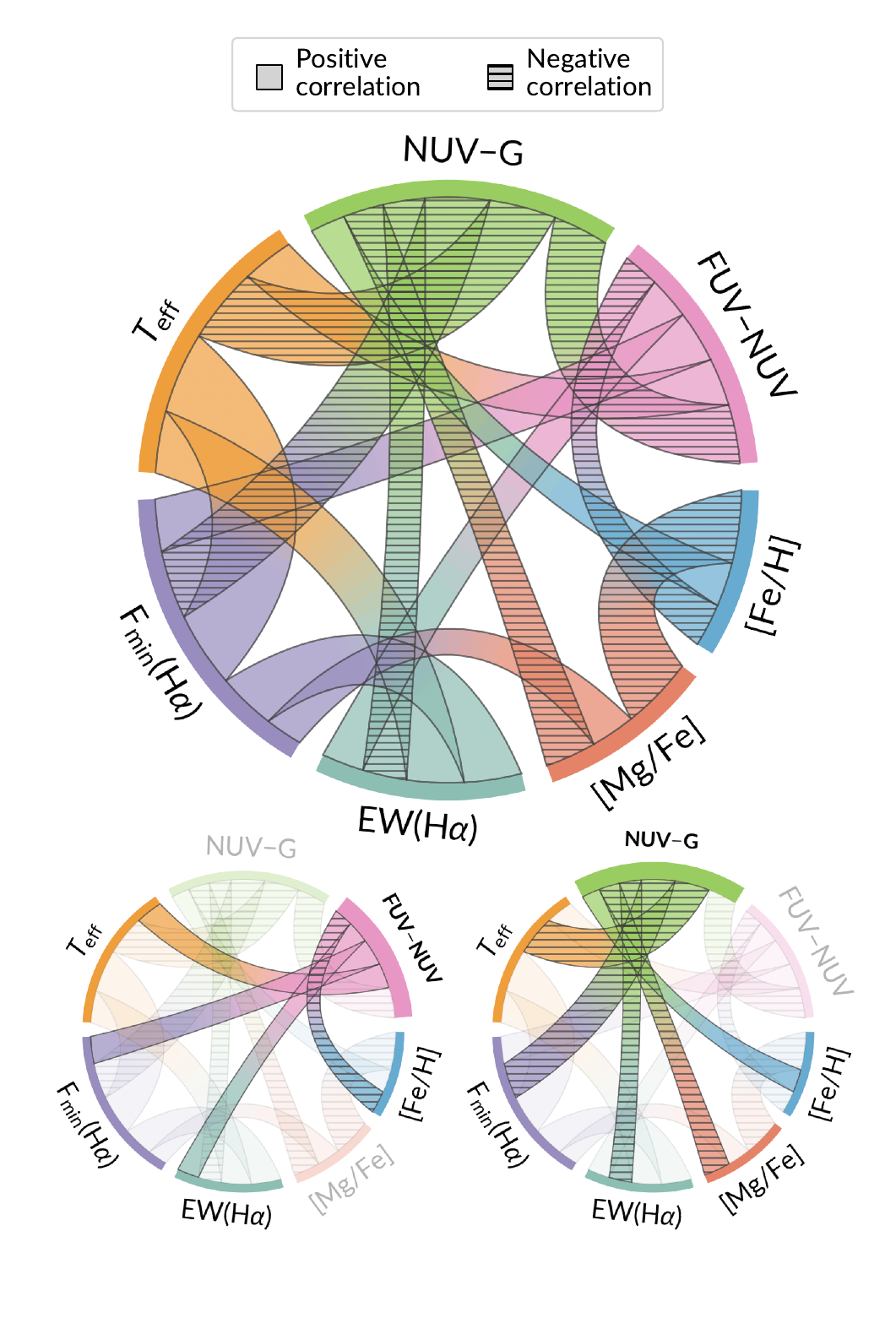}
    \caption{A chord diagram (top) summarising the strongest correlations among stellar parameters in the \tgex\ sample, together with two auxiliary panels (bottom) highlighting the dominant links for the NUV--$G$ and FUV--NUV colours. Chords represent Spearman rank correlations \citep{Spearman1904} with absolute correlation coefficient $|\rho| \geq 0.4$, with colour and hatching indicating the sign of the correlation (positive or negative; see legend). The diagram was generated using \textsc{Cachai} \citep[][but see also \citealt{Gu2014} and \citealt{deSouzaCiardi2015} as references for chord diagrams]{Beltran2025}. The quantities \ewha\ and \fminha\ denote, respectively, the equivalent width and the minimum normalised flux at the core of the H$\alpha$ absorption line, both measured from the normalised spectra. Stellar age is omitted because its correlations fall below the adopted threshold. Correlations involving the H$\alpha$ diagnostics are calculated using only the 24 stars with suitable spectral coverage.}
    \label{fig:chord_diagram}
\end{figure}

Figure \ref{fig:chord_diagram} summarises the strongest correlations among the stellar parameters explored in the \tgex\ sample. Effective temperature (\teff) is a dominant driver, showing strong correlations with the UV--optical colours, especially NUV--$G$, in agreement with the temperature trends identified earlier.

The two H$\alpha$ diagnostics, \ewha\ and \fminha, are tightly correlated with each other and are also linked to \teff\ within the 24-star subsample, reflecting the concentration of elevated H$\alpha$ core flux among its hotter Group~2 stars. Neither diagnostic shows a strong correlation with the UV colours in this subsample. However, correlations involving H$\alpha$ must be interpreted cautiously because the spectral coverage is incomplete and includes only one Group~1 star; they therefore cannot exclude an activity contribution to the UV excess of the broader \tgex\ population.

Chemical abundance parameters form a partially independent substructure. In particular, \mgfe\ correlates with \feh, as expected, while the UV colours remain strongly connected to \teff\ and metallicity. The auxiliary panels further show that NUV--$G$ is the most connected colour in the diagram, whereas FUV--NUV is more selectively linked to \teff, \feh, and \mgfe.

\subsection{Contribution of UV-bright FGK stars to the UV upturn in quiescent galaxies}
\label{subsec:extragalactic}

We conclude the Results section by estimating the contribution of UV-bright FGK stars to the UV emission of quiescent galaxies. Using the empirical UV properties of \tgex\ stars, we quantify the fraction of the observed NUV and FUV emission in UV-upturn galaxies [selected from the GAMA survey \citep{Driver2009} following the selection methods described in \citealt{Dantas2020, Dantas2021}] that can be reproduced as a function of the fraction of FGK stars exhibiting UV-bright behaviour, following the methodology described in Sect. \ref{subsec:method_brute_sed_fit}.

\begin{figure*}
    \centering
    \includegraphics[width=\linewidth]{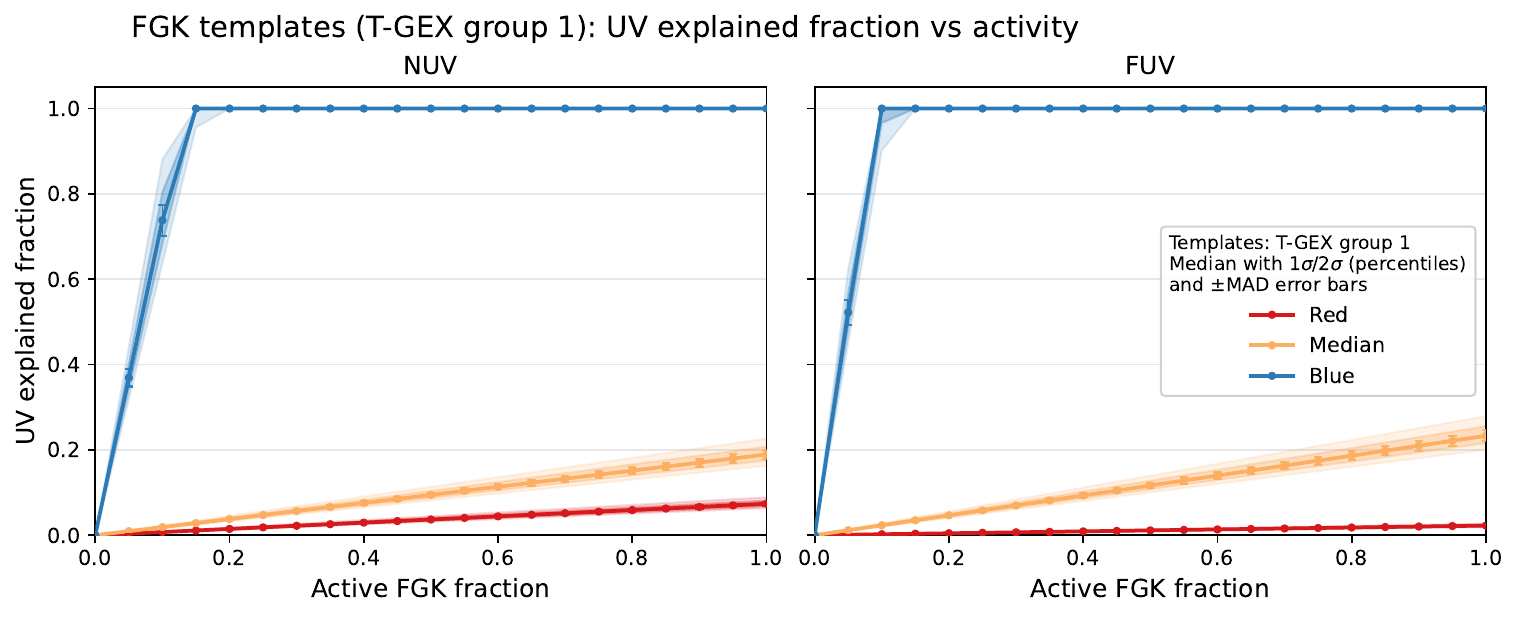}
    \caption{Fraction of the UV emission in quiescent galaxies that can be accounted for by FGK stars with UV excess, estimated with the empirical \tgex-based scaling described in Sect. \ref{subsec:method_brute_sed_fit}. The curves show the median UV-explained fraction across three representative UV-upturn galaxies selected from low-, intermediate-, and high-UV-brightness terciles of the GAMA UV-upturn sample. Results are shown separately for the NUV (left) and FUV (right) as a function of the assumed UV-bright FGK fraction. Coloured curves correspond to three empirical UV templates derived from \tgex\ Group~1: the reddest individual star (red), a median template constructed from the Group~1 median FUV and NUV absolute magnitudes (orange), and the bluest individual star (blue). Solid lines show the median bootstrap estimate; shaded regions indicate the 1$\sigma$ and 2$\sigma$ intervals from 2000 bootstrap realisations that resample galaxy ages, IMF return fractions, and stellar UV luminosities; thin error bars denote the median absolute deviation (MAD).}
    \label{fig:nuv_fuv_contributions}
\end{figure*}

Figure \ref{fig:nuv_fuv_contributions} shows the fraction of the total NUV and FUV emission that can be accounted for under different assumptions for the UV-bright FGK fraction. Here, each `template' refers to an empirical UV luminosity prescription derived from \tgex\ Group 1 and used as a UV building block, analogous to the scaling of stellar components in unresolved simple stellar populations. The reddest and bluest templates correspond to the individual Group 1 stars with the reddest and bluest FUV--NUV colours, respectively, while the median template is constructed from the median FUV and NUV absolute magnitudes of the Group 1 stars. The bluest template has a formal isochrone age estimate of $\sim 200$ Myr, but it lies in a region of the Kiel diagram where age estimates are highly degenerate; we therefore use it only as an empirical UV-luminosity extreme, not as evidence that the corresponding UV-bright phase is necessarily young.

The reddest template contributes negligibly to the FUV emission of UV-upturn galaxies and only weakly to the NUV, while the median template, defined by the UV colour distribution of Group 1, reaches at most $\sim$20\% of the UV output even if all FGK stars are assumed to be UV-bright. The bluest template changes the scaling sharply: under this most UV-luminous empirical case, a UV-excess FGK fraction of only $\sim$10--15\% becomes capable of contributing substantially to the UV budget, provided that such stars are common and long-lived. Even lower fractions ($\lesssim$5\%) can account for $\sim$50\% of the FUV output in this limiting case. These results demonstrate a strongly non-linear dependence on the UV properties of the FGK population, whereby a small admixture of extremely UV-bright stars can dominate the integrated UV emission.

\section{Discussion}
\label{sec:discussion_and_conclusions}

We identified a population of MSTO FGK field stars whose UV properties deviate substantially from the expectations of standard stellar models. Approximately $2/3$ of the \tgex\ sample exhibit UV fluxes in excess of the UV-normal locus defined by \citet{Smith2014}, despite \teff\ and \logg\ consistent with ordinary solar-type stars. The GMM analysis shows that this UV abnormality is structured within the UV--optical colour space, separating the sample into two UV-bright groups and one UV-normal group. These groups also show differences in \teff\ and tentative chemical trends, but given the small and selection-function-limited nature of \tgex, these secondary properties should be interpreted descriptively rather than as intrinsic population signatures. Remarkably, the most extreme group occupies the same region of UV colour space as classical UV-upturn galaxies, despite consisting of apparently low-mass, unevolved stars.

The available H$\alpha$ measurements provide only a preliminary test of chromospheric activity. Among the 24 stars with suitable spectral coverage, enhanced H$\alpha$ core emission is observed only in Group~2 and is associated primarily with hotter and younger UV-normal stars. The observed UV-abnormal stars do not show comparable H$\alpha$ enhancement, suggesting that strong H$\alpha$ activity is not common among the measured UV-excess objects. Nevertheless, only one Group~1 star has H$\alpha$ coverage, and the present data therefore cannot establish whether magnetic or chromospheric activity contributes to the extreme UV behaviour. A more decisive test will require complete and homogeneous coverage using classical activity tracers such as the \ion{Ca}{II} H and K lines, together with additional Balmer and UV diagnostics \citep[][]{Smith2010, Findeisen2011, Linsky1979ApJS...41...47L, Baliunas1995ApJ...438..269B}.

The chemical properties of the UV-abnormal stars should therefore be interpreted cautiously. In the present sample, the UV-abnormal groups occupy narrower metallicity ranges than the UV-normal stars, and the UV-extreme group has a lower median \feh\ and higher median \mgfe. However, given the small size and selection-function-limited nature of \tgex, these trends should not yet be interpreted as intrinsic abundance signatures of UV excess. Instead, they may indicate that UV-abnormal stars preferentially arise in particular Galactic stellar populations, a possibility that requires a dedicated chemo-dynamical analysis.

One important alternative to intrinsic UV emission is unresolved multiplicity or source blending. White-dwarf+FGK binaries are known to mimic this observational configuration: the FGK star can dominate the optical spectrum, while a hot white dwarf or compact companion dominates the UV \citep[][]{Parsons2016}. Similar UV-selected approaches have identified candidate WD+FGK systems in large spectroscopic surveys and nearby-star samples, and UV studies of old stellar populations such as M67 show that evolved or post-interaction binaries can be prominent UV sources \citep[][]{Parsons2016, Nayak2024, Sindhu2018, Reggiani2025}. This possibility is particularly relevant because multiplicity is common among solar-type stars \citep[][]{Offner2023}. We therefore treat unresolved companions, post-interaction products, and chance alignments within the GALEX PSF as important alternatives that cannot be fully excluded with the present data. Our selection explicitly rejects sources with poor \gaia\ astrometric quality, including elevated \texttt{ruwe} and image-parameter diagnostics sensitive to duplicated or irregular astrometric solutions \citep[][as shown in Sect. \ref{sec:data}]{Fabricius2021, Kervella2022}, and also removes spectroscopically peculiar \gaia-ESO sources. These cuts remove many obvious resolved, astrometric, spectroscopic, and peculiar contaminants, but they cannot provide a complete census of close, faint, or UV-dominant companions. Our UV-abnormal stars should therefore be interpreted as candidate cool UV emitters until multi-epoch radial velocities, high-angular-resolution imaging, or UV spectroscopy can establish whether the excess is intrinsic or externally supplied.

The extragalactic implications are therefore significant. Using empirical UV templates derived from \tgex, we find that UV-bright FGK stars can account for a substantial fraction of the UV emission in quiescent UV-upturn galaxies, depending on how extreme the UV-bright tail is and on the fraction of FGK stars that occupy it. For the reddest individual-star template, the contribution is negligible in the FUV and remains small in the NUV even if all FGK stars are UV-bright, while the median Group~1 template reaches at most $\sim$20\%. By contrast, for the bluest individual-star template, the required UV excess FGK fraction falls to $\sim$10--15\%, suggesting that such stars are capable of making a substantial contribution to the observed FUV and NUV output if the UV-bright phase is common and long-lived. This strong non-linearity arises because extreme UV-bright FGK stars sit far above the UV luminosity expected for their mass, allowing a minority population to dominate the integrated signal. These findings do not replace canonical UV-upturn channels, but demonstrate that cool MS stars represent a viable and previously overlooked contributor.

Several caveats apply. Our analysis further assumes a universal Kroupa-like IMF and that the UV-bright FGK stars identified in \tgex\ are representative of the FGK population contributing to the integrated UV emission of quiescent galaxies. Metallicity-dependent UV output, unresolved binarity, residual dust attenuation, and the small, selection-function-limited size of the \tgex\ sample may bias the inferred FGK contribution, although these effects are partially reflected in the bootstrap uncertainty ranges. The \tgex\ sample is small and shaped by the \gaia-ESO selection function, and it excludes M dwarfs, which can be UV-bright through different physical pathways and require separate assessment in stellar populations of varied ages \citep{Shkolnik2011, Shkolnik2014}. In particular, the bluest empirical template should be interpreted as a limiting UV-luminosity case: although its formal isochrone age is young, its position in the Kiel diagram is age-degenerate, and the scaling calculation does not assume that this template necessarily represents a young population.

Our analysis also relies on GALEX photometry, whose $\sim 5.3''$ (NUV) and $\sim 4.2''$ (FUV) spatial resolution means that unresolved UV-bright neighbours, chance alignments, or compact companions within the GALEX PSF cannot be excluded for every source. Moreover, while accretion-related UV production is plausible in general, we find no obvious signatures of active accretion in the available \gaia-ESO wavelength range (e.g. prominent Balmer emission), although weak or intermittent accretion cannot be excluded. More generally, some post-interaction products---such as merger remnants or blue-straggler analogues---could plausibly sustain anomalous UV output and should be considered in future binarity-focused follow-up \citep[][]{Xin2007, Dieball2010, Raso2017, Parsons2016, Offner2023}.

At the same time, the absence of obvious astrometric or spectroscopic peculiarities leaves open the possibility that at least some of these objects reflect an overlooked UV-emitting behaviour of apparently ordinary FGK/MSTO stars. The available H$\alpha$ measurements reveal no strong activity signatures among the observed UV-abnormal stars, but their incomplete coverage, especially in Group~1, prevents this result from being generalised to the full sample. Establishing the duty cycle and physical origin of UV-bright phases in FGK stars will therefore require dedicated observations, particularly multi-epoch radial velocities and high-angular-resolution imaging for binarity and source blending, together with homogeneous optical-activity and blue/UV spectroscopy.

More broadly, this work highlights the value of resolved Galactic field stars as laboratories for interpreting unresolved extragalactic phenomena. A detailed chrono-chemo-dynamical analysis of the \tgex\ stars will provide an important next step toward disentangling the physical origin of their UV behaviour. By directly characterising cool UV emitters in the MW, we motivate a revision of UV population-synthesis models to allow for a non-negligible contribution from cool MS stars, potentially offering a new perspective on the long-standing problem of the UV upturn.

\section{Conclusions}
\label{sec:conc}

We assembled the \tgex\ catalogue of 37 Galactic field FGK/MSTO stars by combining \gaia-ESO spectroscopy with \gaia\ DR3 astrometry, GALEX FUV and NUV photometry, and 2MASS and AllWISE infrared measurements. Astrometric and spectroscopic quality cuts were used to minimise obvious multiplicity and source contamination, although close or UV-dominant companions cannot be excluded. We classified the stellar UV behaviour using an empirical UV-normality criterion and a 3-component GMM, examined H$\alpha$ $\lambda 6563$ activity diagnostics for the 24 stars with suitable coverage, and performed an IMF-based feasibility calculation for quiescent galaxies. Our main conclusions are as follows.

\begin{enumerate}
    \item Approximately 2/3 of the \tgex\ stars are UV-abnormal relative to the empirical FGK UV-normal locus, with the strongest excesses occurring among the coolest stars.

    \item The GMM separates the sample into a cool, strongly UV-excess Group~1 (G1), a hotter and predominantly UV-normal Group~2 (G2), and an intermediate UV-excess Group~3 (G3). G1 and G3 contain only UV-abnormal stars, whereas G2 contains nearly all UV-normal objects.

    \item The groups also differ in their stellar parameters. G1 is the coolest and most $\alpha$-enhanced group, with a narrow, predominantly sub-solar \feh\ distribution. G3 has intermediate temperatures, the highest median \feh, and a broad age distribution. G2 is the hottest group, spans the broadest metallicity range, and has the youngest median age. These trends remain descriptive because of the small and selection-function-limited sample.

    \item Within the H$\alpha$ subsample, enhanced core emission is found only among hotter G2 stars. None of the observed UV-abnormal stars shows comparable enhancement, but the presence of only one G1 star with H$\alpha$ coverage prevents a representative test of chromospheric activity among the UV-extreme stars.

    \item The extragalactic scaling is strongly template-dependent. The median G1 template accounts for at most $\sim20\%$ of the observed UV output, whereas the bluest limiting template could match most or all of the NUV and FUV emission if shared by $\sim10$--$15\%$ of surviving FGK stars. This demonstrates feasibility rather than a decomposition of the UV upturn, and unresolved companions, post-interaction products, and blending remain viable origins of the stellar UV excess.
\end{enumerate}

The \tgex\ systems should therefore be regarded as candidate cool UV emitters and as a potentially important additional channel rather than a replacement for canonical UV-upturn sources. Establishing whether their UV emission is intrinsic, companion-driven, or related to stellar interaction will require homogeneous activity diagnostics, multi-epoch radial velocities, high-angular-resolution imaging, and UV spectroscopy. The details on the chrono-chemo-dynamical properties of the \tgex\ sample will be explored in \citet[][\citetalias{Beltran2026_TGEXII} of this series]{Beltran2026_TGEXII}.

\begin{acknowledgements}
MLLD acknowledges the support of Agencia Nacional de Investigación y Desarrollo (ANID), Chile, through Fondecyt Postdoctorado Folio 3240344. MLLD is grateful for Miúcha, her feline companion, for the long-standing companionship, love, and support.
MLLD and DB acknowledge support from the Pontificia Universidad Católica de Chile for the summer program ``IPRE''. 
MLLD, DB, and PBT acknowledge ANID Basal Project FB210003. 
REG thanks INAF for the support (Large Grants EPOCH and WST), the Mini-Grants Checs (1.05.23.04.02), and the financial support under the National Recovery and Resilience Plan (NRRP), Mission 4, Component 2, Investment 1.1, Call for tender No. 104 published on 2.2.2022 by the Italian Ministry of University and Research (MUR), funded by the European Union – NextGenerationEU – Project ‘Cosmic POT’ (PI: L. Magrini) Grant Assignment Decree No. 2022X4TM3H by the Italian Ministry of the University and Research (MUR).
RS acknowledges support from the National Science Centre, Poland, project 2019/34/E/ST9/00133.
PC acknowledges support from Conselho Nacional de Desenvolvimento Cient\'ifico e Tecnol\'ogico (CNPq) under grant 308715/2025-0.
PBT acknowledges partial support from Fondecyt Regular 1240465.
\end{acknowledgements}

%

\bibliographystyle{bibtex/aa}          
\bibliography{paper.bib}        

@article{Beltran2026_TGEXII,
    author  = {Beltr{\'a}n, D. and Dantas, M. L. L. and Smiljanic, R. and Tissera, P. B.},
    title   = {The {TGEX} project. II. Chemo-dynamics of {UV}-bright {FGK} stars: $\alpha$-enhancement and hypernova-like fingerprints},
    journal = {\aap},
    year    = {2026},
    note    = {in preparation}
}

@ARTICLE{Bahcall1983,
    author = {{Bahcall}, J.~N. and {Piran}, T.},
    title = "{Stellar collapses in the Galaxy.}",
    journal = {\apjl},
    year = 1983,
    month = apr,
    volume = {267},
    pages = {L77-L81},
    doi = {10.1086/184007},
    adsurl = {https://ui.adsabs.harvard.edu/abs/1983ApJ...267L..77B}
}

@ARTICLE{Baldry2018,
    author = {{Baldry}, I.~K. and {Liske}, J. and {Brown}, M.~J.~I. and {Robotham}, A.~S.~G. and {Driver}, S.~P. and {Dunne}, L. and {Alpaslan}, M. and {Brough}, S. and {Cluver}, M.~E. and {Eardley}, E. and {Farrow}, D.~J. and {Heymans}, C. and 
	{Hildebrandt}, H. and {Hopkins}, A.~M. and {Kelvin}, L.~S. and {Loveday}, J. and {Moffett}, A.~J. and {Norberg}, P. and {Owers}, M.~S. and {Taylor}, E.~N. and {Wright}, A.~H. and {Bamford}, S.~P. and {Bland-Hawthorn}, J. and {Bourne}, N. and {Bremer}, M.~N. and {Colless}, M. and {Conselice}, C.~J. and {Croom}, S.~M. and {Davies}, L.~J.~M. and {Foster}, C. and {Grootes}, M.~W. and {Holwerda}, B.~W. and {Jones}, D.~H. and {Kafle}, P.~R. and {Kuijken}, K. and {Lara-Lopez}, M.~A. and {L{\'o}pez-S{\'a}nchez}, {\'A}.~R. and 
	{Meyer}, M.~J. and {Phillipps}, S. and {Sutherland}, W.~J. and {van Kampen}, E. and {Wilkins}, S.~M.},
    title = "{Galaxy And Mass Assembly: the G02 field, Herschel-ATLAS target selection and data release 3}",
    journal = {\mnras},
    year = 2018,
    month = mar,
    volume = 474,
    pages = {3875-3888},
    doi = {10.1093/mnras/stx3042},
}

@ARTICLE{Baliunas1995ApJ...438..269B,
    author = {{Baliunas}, S.~L. and {Donahue}, R.~A. and {Soon}, W.~H. and {Horne}, J.~H. and {Frazer}, J. and {Woodard-Eklund}, L. and {Bradford}, M. and {Rao}, L.~M. and {Wilson}, O.~C. and {Zhang}, Q. and {Bennett}, W. and {Briggs}, J. and {Carroll}, S.~M. and {Duncan}, D.~K. and {Figueroa}, D. and {Lanning}, H.~H. and {Misch}, T. and {Mueller}, J. and {Noyes}, R.~W. and {Poppe}, D. and {Porter}, A.~C. and {Robinson}, C.~R. and {Russell}, J. and {Shelton}, J.~C. and {Soyumer}, T. and {Vaughan}, A.~H. and {Whitney}, J.~H.},
    title = "{Chromospheric Variations in Main-Sequence Stars. II.}",
    journal = {\apj},
    year = 1995,
    month = jan,
    volume = {438},
    pages = {269},
    doi = {10.1086/175072},
    adsurl = {https://ui.adsabs.harvard.edu/abs/1995ApJ...438..269B}
}

@ARTICLE{Beltran2025,
    author = {{Beltr{\'a}n}, D. and {Dantas}, M.~L.~L.},
    title = "{CACHAI's First Module: A Fully Customizable Chord Diagram for Astronomy and Beyond}",
    journal = {Research Notes of the American Astronomical Society},
    year = 2025,
    month = aug,
    volume = {9},
    number = {8},
    eid = {216},
    pages = {216},
    doi = {10.3847/2515-5172/adf8df},
    adsurl = {https://ui.adsabs.harvard.edu/abs/2025RNAAS...9..216B}
}

@ARTICLE{BertinArnouts1996,
    author = {{Bertin}, E. and {Arnouts}, S.},
    title = "{SExtractor: Software for source extraction.}",
    journal = {\aaps},
    year = 1996,
    month = jun,
    volume = {117},
    pages = {393-404},
    doi = {10.1051/aas:1996164},
    adsurl = {https://ui.adsabs.harvard.edu/abs/1996A&AS..117..393B}
}

@ARTICLE{Bertola1982,
    author = {{Bertola}, F. and {Capaccioli}, M. and {Oke}, J.~B.},
    title = "{IUE observations of NGC 4649, an elliptical galaxy with a strong ultraviolet flux.}",
    journal = {\apj},
    year = 1982,
    month = mar,
    volume = {254},
    pages = {494-499},
    doi = {10.1086/159758}
}

@ARTICLE{BHG2016,
    author = {{Bland-Hawthorn}, Joss and {Gerhard}, Ortwin},
    title = "{The Galaxy in Context: Structural, Kinematic, and Integrated Properties}",
    journal = {\araa},
    year = 2016,
    month = sep,
    volume = {54},
    pages = {529-596},
    doi = {10.1146/annurev-astro-081915-023441}
}

@article{Bianchi2014,
    author = {Bianchi, Luciana},
    doi = {10.1007/s10509-014-1935-6},
    issn = {1572946X},
    journal = {Astrophysics and Space Science},
    number = {1},
    pages = {103--112},
    title = {{The Galaxy Evolution Explorer (GALEX). Its legacy of UV surveys, and science highlights}},
    volume = {354},
    year = {2014}
}

@ARTICLE{Bressan2012,
    author = {{Bressan}, Alessandro and {Marigo}, Paola and {Girardi}, L{\'e}o. and {Salasnich}, Bernardo and {Dal Cero}, Claudia and {Rubele}, Stefano and {Nanni}, Ambra},
    title = "{PARSEC: stellar tracks and isochrones with the PAdova and TRieste Stellar Evolution Code}",
    journal = {\mnras},
    year = 2012,
    month = nov,
    volume = {427},
    number = {1},
    pages = {127-145},
    doi = {10.1111/j.1365-2966.2012.21948.x}
}

@ARTICLE{Brown2000,
    author = {{Brown}, T.~M. and {Bowers}, C.~W. and {Kimble}, R.~A. and {Sweigart}, A.~V. and 
	{Ferguson}, H.~C.},
    title = "{Detection and Photometry of Hot Horizontal Branch Stars in the Core of M32}",
    journal = {\apj},
    year = 2000,
    month = mar,
    volume = 532,
    pages = {308-322},
    doi = {10.1086/308566}
}

@ARTICLE{Burstein1988,
    author = {{Burstein}, David and {Bertola}, F. and {Buson}, L.~M. and {Faber}, S.~M. and {Lauer}, Tod R.},
    title = "{The Far-Ultraviolet Spectra of Early-Type Galaxies}",
    journal = {\apj},
    year = 1988,
    month = May,
    volume = {328},
    pages = {440},
    doi = {10.1086/166304}
}

@ARTICLE{Buson2006,
    author = {{Buson}, L.~M. and {Bertone}, E. and {Buzzoni}, A. and {Carraro}, G.},
    title = "{New Candidate EHB Stars in the Open Cluster NGC 6791: Looking Locally Into the UV-Upturn Phenomenon}",
    journal = {Baltic Astronomy},
    year = 2006,
    month = jan,
    volume = {15},
    pages = {49-52},
    archivePrefix = {arXiv},
    eprint = {astro-ph/0509772},
    primaryClass = {astro-ph},
}

@ARTICLE{Buzzoni2012,
    author = {{Buzzoni}, Alberto and {Bertone}, Emanuele and {Carraro}, Giovanni and {Buson}, Lucio},
    title = "{Stellar Lifetime and Ultraviolet Properties of the Old Metal-rich Galactic Open Cluster NGC 6791: A Pathway to Understand the Ultraviolet Upturn of Elliptical Galaxies}",
    journal = {\apj},
    year = 2012,
    month = apr,
    volume = {749},
    number = {1},
    eid = {35},
    pages = {35},
    doi = {10.1088/0004-637X/749/1/35},
}

@ARTICLE{Chabrier2003,
    author = {{Chabrier}, Gilles},
    title = "{Galactic Stellar and Substellar Initial Mass Function}",
    journal = {\pasp},
    year = 2003,
    month = jul,
    volume = {115},
    number = {809},
    pages = {763-795},
    doi = {10.1086/376392},
    archivePrefix = {arXiv},
    eprint = {astro-ph/0304382},
    primaryClass = {astro-ph},
    adsurl = {https://ui.adsabs.harvard.edu/abs/2003PASP..115..763C}
}

@ARTICLE{CodeWelch1979,
    author = {{Code}, A.~D. and {Welch}, G.~A.},
    title = "{Ultraviolet photometry from the Orbiting Astronomical Observatory. XXVI. Energy distributions of seven early-type galaxies and the central bulge of M31.}",
    journal = {\apj},
    year = 1979,
    month = feb,
    volume = {228},
    pages = {95-104},
    doi = {10.1086/156825}
}

@ARTICLE{Conroy2013,
    author = {{Conroy}, Charlie},
    title = "{Modeling the Panchromatic Spectral Energy Distributions of Galaxies}",
    journal = {\araa},
    year = 2013,
    month = aug,
    volume = {51},
    number = {1},
    pages = {393-455},
    doi = {10.1146/annurev-astro-082812-141017}
}

@ARTICLE{Cretignier2020,
    author = {{Cretignier}, M. and {Francfort}, J. and {Dumusque}, X. and {Allart}, R. and {Pepe}, F.},
    title = "{RASSINE: Interactive tool for normalising stellar spectra. I. Description and performance of the code}",
    journal = {\aap},
    year = 2020,
    month = aug,
    volume = {640},
    eid = {A42},
    pages = {A42},
    doi = {10.1051/0004-6361/202037722}
}

@ARTICLE{Cutri2014,
   author = {{Cutri}, R.~M. and {Wright}, E.~L. and {Conrow}, T. and {Fowler}, J.~W. and {Eisenhardt}, P.~R.~M. and {Grillmair}, C. and {Kirkpatrick}, J.~D. and {Masci}, F. and {McCallon}, H.~L. and {Wheelock}, S.~L. and {Fajardo-Acosta}, S. and {Yan}, L. and {Benford}, D. and {Harbut}, M. and {Jarrett}, T. and {Lake}, S. and {Leisawitz}, D. and {Ressler}, M.~E. and {Stanford}, S.~A. and {Tsai}, C. -W. and {Liu}, F. and {Helou}, G. and {Mainzer}, A. and {Gettngs}, D. and {Gonzalez}, A. and {Hoffman}, D. and {Marsh}, K.~A. and {Padgett}, D. and {Skrutskie}, M.~F. and {Beck}, R. and {Papin}, M. and {Wittman}, M.},
   title = "{VizieR Online Data Catalog: AllWISE Data Release (Cutri+ 2013)}",
   journal = {VizieR Online Data Catalog},
   year = 2014,
   month = feb,
   eid = {II/328},
   pages = {II/328},
   adsurl = {https://ui.adsabs.harvard.edu/abs/2014yCat.2328....0C}
}

@ARTICLE{Dantas2020,
    author = {{Dantas}, M.~L.~L. and {Coelho}, P.~R.~T. and {de Souza}, R.~S. and {Gon{\c{c}}alves}, T.~S.},
    title = "{UV bright red-sequence galaxies: how do UV upturn systems evolve in redshift and stellar mass?}",
    journal = {\mnras},
    year = 2020,
    month = feb,
    volume = {492},
    number = {2},
    pages = {2996-3011},
    doi = {10.1093/mnras/stz3609}
}

@ARTICLE{Dantas2021,
    author = {{Dantas}, M.~L.~L. and {Coelho}, P.~R.~T. and {S{\'a}nchez-Bl{\'a}zquez}, P.},
    title = "{UV upturn versus UV weak galaxies: differences and similarities of their stellar populations unveiled by a de-biased sample}",
    journal = {\mnras},
    year = 2021,
    month = jan,
    volume = {500},
    number = {2},
    pages = {1870-1883},
    doi = {10.1093/mnras/staa3447}
}

@ARTICLE{Dantas2023,
    author = {{Dantas}, M.~L.~L. and {Smiljanic}, R. and {Boesso}, R. and {Rocha-Pinto}, H.~J. and {Magrini}, L. and {Guiglion}, G. and {Tautvai{\v{s}}ien{\.{e}}}, G. and {Gilmore}, G. and {Randich}, S. and {Bensby}, T. and {Bragaglia}, A. and {Bergemann}, M. and {Carraro}, G. and {Jofr{\'e}}, P. and {Zaggia}, S.},
    title = "{The Gaia-ESO Survey: Old super-metal-rich visitors from the inner Galaxy}",
    journal = {\aap},
    year = 2023,
    month = jan,
    volume = {669},
    eid = {A96},
    pages = {A96},
    doi = {10.1051/0004-6361/202243667}
}

@article{Dantas2025a,
    author = {{Dantas}, M. L. L. and {Smiljanic}, R. and {de Souza}, R. S. and {Tissera}, P. B. and {Magrini}, L.},
    title = {Probing the origins. I. Generalised Additive Model inference of birth radii for Milky Way stars in the solar vicinity},
    journal = {\aap},
    volume = {696},
    eid = {A205},
    pages = {A205},
    year = {2025},
    month = apr,
    doi = {10.1051/0004-6361/202453034},
    url = {https://doi.org/10.1051/0004-6361/202453034},
    sortkey={001}
}

@ARTICLE{deLaverny2025,
    author = {{de Laverny}, Patrick and {Ligi}, Roxanne and {Crida}, Aur{\'e}lien and {Recio-Blanco}, Alejandra and {Palicio}, Pedro A.},
    title = "{The Gaia spectroscopic catalogue of exoplanets and host stars}",
    journal = {\aap},
    year = 2025,
    month = jul,
    volume = {699},
    eid = {A100},
    pages = {A100},
    doi = {10.1051/0004-6361/202554739}
}

@ARTICLE{deSouzaCiardi2015,
    author = {{de Souza}, R.~S. and {Ciardi}, B.},
    title = "{AMADA-Analysis of multidimensional astronomical datasets}",
    journal = {Astronomy and Computing},
    year = 2015,
    month = sep,
    volume = {12},
    pages = {100-108},
    doi = {10.1016/j.ascom.2015.06.006},
    archivePrefix = {arXiv},
    eprint = {1503.07736},
    primaryClass = {astro-ph.IM},
    adsurl = {https://ui.adsabs.harvard.edu/abs/2015A&C....12..100D}
}

@ARTICLE{deSouza2017,
    author = {{de Souza}, R.~S. and {Dantas}, M.~L.~L. and {Costa-Duarte}, M.~V. and {Feigelson}, E.~D. and {Killedar}, M. and {Lablanche}, P.-Y. and {Vilalta}, R. and {Krone-Martins}, A. and {Beck}, R. and {Gieseke}, F.},
    title = "{A probabilistic approach to emission-line galaxy classification}",
    journal = {\mnras},
    year = 2017,
    month = dec,
    volume = {472},
    number = {3},
    pages = {2808-2822},
    doi = {10.1093/mnras/stx2156},
    archivePrefix = {arXiv},
    eprint = {1703.07607},
    primaryClass = {astro-ph.GA},
    adsurl = {https://ui.adsabs.harvard.edu/abs/2017MNRAS.472.2808D}
}

@ARTICLE{Dieball2010,
    author = {{Dieball}, Andrea and {Long}, Knox S. and {Knigge}, Christian and {Thomson}, Grace S. and {Zurek}, David R.},
    title = "{A Far-ultraviolet Survey of M80: X-Ray Source Counterparts, Strange Blue Stragglers, and the Recovery of Nova T Sco}",
    journal = {\apj},
    year = 2010,
    month = feb,
    volume = {710},
    number = {1},
    pages = {332-345},
    doi = {10.1088/0004-637X/710/1/332}
}

@ARTICLE{Dorman1993,
    author = {{Dorman}, Ben and {Rood}, Robert T. and {O'Connell}, Robert W.},
    title = "{Ultraviolet Radiation from Evolved Stellar Populations. I. Models}",
    journal = {\apj},
    year = 1993,
    month = dec,
    volume = {419},
    pages = {596},
    doi = {10.1086/173511}
}

@ARTICLE{Driver2009,
    author = {{Driver}, Simon P. and {Norberg}, Peder and {Baldry}, Ivan K. and {Bamford}, Steven P. and {Hopkins}, Andrew M. and {Liske}, Jochen and {Loveday}, Jon and {Peacock}, John A. and {Hill}, D.~T. and {Kelvin}, L.~S. and {Robotham}, A.~S.~G. and {Cross}, N.~J.~G. and {Parkinson}, H.~R. and {Prescott}, M. and {Conselice}, C.~J. and {Dunne}, L. and {Brough}, S. and {Jones}, H. and {Sharp}, R.~G. and {van Kampen}, E. and {Oliver}, S. and {Roseboom}, I.~G. and {Bland-Hawthorn}, J. and {Croom}, S.~M. and {Ellis}, S. and {Cameron}, E. and {Cole}, S. and {Frenk}, C.~S. and {Couch}, W.~J. and {Graham}, A.~W. and {Proctor}, R. and {De Propris}, R. and {Doyle}, I.~F. and {Edmondson}, E.~M. and {Nichol}, R.~C. and {Thomas}, D. and {Eales}, S.~A. and {Jarvis}, M.~J. and {Kuijken}, K. and {Lahav}, O. and {Madore}, B.~F. and {Seibert}, M. and {Meyer}, M.~J. and {Staveley-Smith}, L. and {Phillipps}, S. and {Popescu}, C.~C. and {Sansom}, A.~E. and {Sutherland}, W.~J. and {Tuffs}, R.~J. and {Warren}, S.~J.},
    title = "{GAMA: towards a physical understanding of galaxy formation}",
    journal = {Astronomy and Geophysics},
    year = 2009,
    month = oct,
    volume = {50},
    number = {5},
    pages = {5.12-5.19},
    doi = {10.1111/j.1468-4004.2009.50512.x},
    archivePrefix = {arXiv},
    eprint = {0910.5123},
    primaryClass = {astro-ph.CO},
    adsurl = {https://ui.adsabs.harvard.edu/abs/2009A&G....50e..12D}
}

@ARTICLE{Eldridge2017,
    author = {{Eldridge}, J.~J. and {Stanway}, E.~R. and {Xiao}, L. and {McClelland}, L.~A.~S. and {Taylor}, G. and {Ng}, M. and {Greis}, S.~M.~L. and {Bray}, J.~C.},
    title = "{Binary Population and Spectral Synthesis Version 2.1: Construction, Observational Verification, and New Results}",
    journal = {\pasa},
    year = 2017,
    month = nov,
    volume = {34},
    eid = {e058},
    pages = {e058},
    doi = {10.1017/pasa.2017.51}
}

@ARTICLE{Fabricius2021,
    author = {{Fabricius}, C. and {Luri}, X. and {Arenou}, F. and {Babusiaux}, C. and {Helmi}, A. and {Muraveva}, T. and {Reyl{\'e}}, C. and {Spoto}, F. and {Vallenari}, A. and {Antoja}, T. and {Balbinot}, E. and {Barache}, C. and {Bauchet}, N. and {Bragaglia}, A. and {Busonero}, D. and {Cantat-Gaudin}, T. and {Carrasco}, J.~M. and {Diakit{\'e}}, S. and {Fabrizio}, M. and {Figueras}, F. and {Garcia-Gutierrez}, A. and {Garofalo}, A. and {Jordi}, C. and {Kervella}, P. and {Khanna}, S. and {Leclerc}, N. and {Licata}, E. and {Lambert}, S. and {Marrese}, P.~M. and {Masip}, A. and {Ramos}, P. and {Robichon}, N. and {Robin}, A.~C. and {Romero-G{\'o}mez}, M. and {Rubele}, S. and {Weiler}, M.},
    title = "{Gaia Early Data Release 3. Catalogue validation}",
    journal = {\aap},
    year = 2021,
    month = may,
    volume = {649},
    eid = {A5},
    pages = {A5},
    doi = {10.1051/0004-6361/202039834}
}

@ARTICLE{Findeisen2011,
    author = {{Findeisen}, K. and {Hillenbrand}, L. and {Soderblom}, D.},
    title = "{Stellar Activity in the Broadband Ultraviolet}",
    journal = {\aj},
    year = 2011,
    month = jul,
    volume = {142},
    number = {1},
    eid = {23},
    pages = {23},
    doi = {10.1088/0004-6256/142/1/23}
}

@ARTICLE{FBH2002,
    author = {{Freeman}, Ken and {Bland-Hawthorn}, Joss},
    title = "{The New Galaxy: Signatures of Its Formation}",
    journal = {\araa},
    year = 2002,
    month = jan,
    volume = {40},
    pages = {487-537},
    doi = {10.1146/annurev.astro.40.060401.093840}
}

@ARTICLE{Gaia2023_DR3,
    author = {{Gaia Collaboration} and {Vallenari}, A. and {Brown}, A.~G.~A. and {Prusti}, T. and {de Bruijne}, J.~H.~J. and {Arenou}, F. and {Babusiaux}, C. and {Biermann}, M. and others},
    title = "{Gaia Data Release 3. Summary of the content and survey properties}",
    journal = {\aap},
    year = 2023,
    month = jun,
    volume = {674},
    eid = {A1},
    pages = {A1},
    doi = {10.1051/0004-6361/202243940},
    archivePrefix = {arXiv},
    eprint = {2208.00211},
    primaryClass = {astro-ph.GA},
    adsurl = {https://ui.adsabs.harvard.edu/abs/2023A&A...674A...1G}
}

@ARTICLE{GarciaDelgado2026,
    author = {{Garc{\'\i}a-Delgado}, Juan Jos{\'e} and {Dantas}, M.~L.~L. and {Rebollido}, Isabel and {Smiljanic}, Rodolfo},
    title = "{Probing the origins. III. Exoplanet demographics across Galactic birth radii}",
    journal = {arXiv e-prints},
    year = 2026,
    month = jul,
    eid = {arXiv:2607.07787},
    pages = {arXiv:2607.07787},
    doi = {10.48550/arXiv.2607.07787},
    archivePrefix = {arXiv},
    eprint = {2607.07787},
    primaryClass = {astro-ph.GA},
    adsurl = {https://ui.adsabs.harvard.edu/abs/2026arXiv260707787J}
}

@ARTICLE{Gilmore2022,
    author = {{Gilmore}, G. and {Randich}, S. and {Worley}, C.~C. and {Hourihane}, A. and others},
    title = "{The Gaia-ESO Public Spectroscopic Survey: Motivation, implementation, GIRAFFE data processing, analysis, and final data products}",
    journal = {\aap},
    year = 2022,
    month = oct,
    volume = {666},
    eid = {A120},
    pages = {A120},
    doi = {10.1051/0004-6361/202243134}
}

@ARTICLE{Goncalves2020,
    author = {{Gon{\c{c}}alves}, Geraldo and {Coelho}, Paula and {Schiavon}, Ricardo and {Usher}, Christopher},
    title = "{How well can we determine ages and chemical abundances from spectral fitting of integrated light spectra?}",
    journal = {\mnras},
    year = 2020,
    month = dec,
    volume = {499},
    number = {2},
    pages = {2327-2339},
    doi = {10.1093/mnras/staa3051},
    archivePrefix = {arXiv},
    eprint = {2010.02940},
    primaryClass = {astro-ph.GA},
    adsurl = {https://ui.adsabs.harvard.edu/abs/2020MNRAS.499.2327G}
}

@ARTICLE{Goudfrooij2018,
    author = {{Goudfrooij}, Paul},
    title = "{Dissolved Massive Metal-rich Globular Clusters Can Cause the Range of UV Upturn Strengths Found among Early-type Galaxies}",
    journal = {\apj},
    year = 2018,
    month = apr,
    volume = {857},
    number = {1},
    eid = {16},
    pages = {16},
    doi = {10.3847/1538-4357/aab553}
}

@ARTICLE{Green2018,
    author = {{Green}, Gregory M. and {Schlafly}, Edward F. and {Finkbeiner}, Douglas and {Rix}, Hans-Walter and {Martin}, Nicolas and {Burgett}, William and {Draper}, Peter W. and {Flewelling}, Heather and {Hodapp}, Klaus and {Kaiser}, Nicholas and {Kudritzki}, Rolf-Peter and {Magnier}, Eugene A. and {Metcalfe}, Nigel and {Tonry}, John L. and {Wainscoat}, Richard and {Waters}, Christopher},
    title = "{Galactic reddening in 3D from stellar photometry - an improved map}",
    journal = {\mnras},
    year = 2018,
    month = jul,
    volume = {478},
    number = {1},
    pages = {651-666},
    doi = {10.1093/mnras/sty1008}
}

@ARTICLE{Green2018_dustmaps,
    author = {{Green}, {Gregory M.}},
    title = "{dustmaps: A Python interface for maps of interstellar dust}",
    journal = {The Journal of Open Source Software},
    year = "2018",
    month = "Jun",
    volume = {3},
    number = {26},
    pages = {695},
    doi = {10.21105/joss.00695},
    adsurl = {https://ui.adsabs.harvard.edu/abs/2018JOSS....3..695M}
}

@article{Gu2014,
    title = {circlize implements and enhances circular visualization in R},
    volume = {30},
    ISSN = {1367-4803},
    url = {http://dx.doi.org/10.1093/bioinformatics/btu393},
    DOI = {10.1093/bioinformatics/btu393},
    number = {19},
    journal = {Bioinformatics},
    publisher = {Oxford University Press (OUP)},
    author = {Gu,  Zuguang and Gu,  Lei and Eils,  Roland and Schlesner,  Matthias and Brors,  Benedikt},
    year = {2014},
    month = jun,
    pages = {2811–2812}
}

@ARTICLE{Han2007,
    author = {{Han}, Z. and {Podsiadlowski}, Ph. and {Lynas-Gray}, A.~E.},
    title = "{A binary model for the UV-upturn of elliptical galaxies}",
    journal = {\mnras},
    year = 2007,
    month = sep,
    volume = {380},
    number = {3},
    pages = {1098-1118},
    doi = {10.1111/j.1365-2966.2007.12151.x}
}

@ARTICLE{Hardorp1978,
    author = {{Hardorp}, J.},
    title = "{The sun among the stars. I. A search for solar spectral analogs.}",
    journal = {\aap},
    year = 1978,
    month = feb,
    volume = {63},
    pages = {383-390},
    adsurl = {https://ui.adsabs.harvard.edu/abs/1978A&A....63..383H}
}

@ARTICLE{Hardorp1982,
    author = {{Hardorp}, J.},
    title = "{The sun among the stars. V. A second search for solar spectral analogs. The Hyades' distance.}",
    journal = {\aap},
    year = 1982,
    month = jan,
    volume = {105},
    pages = {120-132},
    adsurl = {https://ui.adsabs.harvard.edu/abs/1982A&A...105..120H}
}

@book{Hastie2001,
    address = {New York, NY, USA},
    author = {Hastie, Trevor and Tibshirani, Robert and Friedman, Jerome},
    publisher = {Springer New York Inc.},
    series = {Springer Series in Statistics},
    title = {The Elements of Statistical Learning},
    year = 2001
}

@ARTICLE{Herczeg2008,
    author = {{Herczeg}, Gregory J. and {Hillenbrand}, Lynne A.},
    title = "{UV Excess Measures of Accretion onto Young Very Low Mass Stars and Brown Dwarfs}",
    journal = {\apj},
    year = 2008,
    month = jul,
    volume = {681},
    number = {1},
    pages = {594-625},
    doi = {10.1086/586728},
    archivePrefix = {arXiv},
    eprint = {0801.3525},
    primaryClass = {astro-ph},
    adsurl = {https://ui.adsabs.harvard.edu/abs/2008ApJ...681..594H}
}

@ARTICLE{Helmi2020,
    author = {{Helmi}, Amina},
    title = "{Streams, Substructures, and the Early History of the Milky Way}",
    journal = {\araa},
    year = 2020,
    month = aug,
    volume = {58},
    pages = {205-256},
    doi = {10.1146/annurev-astro-032620-021917}
}

@article{HPB2013,
    author = {Hernández-Pérez, Fabiola and Bruzual, Gustavo},
    title = "{Revisiting binary stars in population synthesis models}",
    journal = {\mnras},
    volume = {431},
    number = {3},
    pages = {2612-2621},
    year = {2013},
    month = {03}
}

@ARTICLE{HPB2014,
    author = {{Hern{\'a}ndez-P{\'e}rez}, Fabiola and {Bruzual}, Gustavo},
    title = "{Binary stars and the UVX in early-type galaxies}",
    journal = {\mnras},
    year = 2014,
    month = nov,
    volume = {444},
    number = {3},
    pages = {2571-2579},
    doi = {10.1093/mnras/stu1627}
}

@ARTICLE{HBH2021,
    author = {{Hui-Bon-Hoa}, A.},
    title = "{Stellar models with self-consistent Rosseland opacities. Consequences for stellar structure and evolution}",
    journal = {\aap},
    year = 2021,
    month = feb,
    volume = {646},
    eid = {L6},
    pages = {L6},
    doi = {10.1051/0004-6361/202040095},
    adsurl = {https://ui.adsabs.harvard.edu/abs/2021A&A...646L...6H},
vadsnote = {Provided by the SAO/NASA Astrophysics Data System}
}

@ARTICLE{Iglesias1996,
    author = {{Iglesias}, Carlos A. and {Rogers}, Forrest J.},
    title = "{Updated Opal Opacities}",
    journal = {\apj},
    year = 1996,
    month = jun,
    volume = {464},
    pages = {943},
    doi = {10.1086/177381},
    adsurl = {https://ui.adsabs.harvard.edu/abs/1996ApJ...464..943I}
}

@book{Hansen2004,
    author = {{Hansen}, Carl J. and {Kawaler}, Steven D. and {Trimble},  Virginia},
    title = {Stellar Interiors: Physical Principles,  Structure,  and Evolution},
    ISBN = {9781441991102},
    ISSN = {0941-7834},
    DOI = {10.1007/978-1-4419-9110-2},
    year = {2004}
}

@ARTICLE{Hurley2000,
    author = {{Hurley}, Jarrod R. and {Pols}, Onno R. and {Tout}, Christopher A.},
    title = "{Comprehensive analytic formulae for stellar evolution as a function of mass and metallicity}",
    journal = {\mnras},
    year = 2000,
    month = jul,
    volume = {315},
    number = {3},
    pages = {543-569},
    doi = {10.1046/j.1365-8711.2000.03426.x}
}

@ARTICLE{Kervella2022,
    author = {{Kervella}, Pierre and {Arenou}, Fr{\'e}d{\'e}ric and {Th{\'e}venin}, Fr{\'e}d{\'e}ric},
    title = "{Stellar and substellar companions from Gaia EDR3. Proper-motion anomaly and resolved common proper-motion pairs}",
    journal = {\aap},
    year = 2022,
    month = jan,
    volume = {657},
    eid = {A7},
    pages = {A7},
    doi = {10.1051/0004-6361/202142146},
    archivePrefix = {arXiv},
    eprint = {2109.10912},
    primaryClass = {astro-ph.SR},
    adsurl = {https://ui.adsabs.harvard.edu/abs/2022A&A...657A...7K}
}

@ARTICLE{Kroupa2001,
    author = {{Kroupa}, Pavel},
    title = "{On the variation of the initial mass function}",
    journal = {\mnras},
    year = 2001,
    month = apr,
    volume = {322},
    number = {2},
    pages = {231-246},
    doi = {10.1046/j.1365-8711.2001.04022.x}
}

@ARTICLE{Leitner2011,
    author = {{Leitner}, Samuel N. and {Kravtsov}, Andrey V.},
    title = "{Fuel Efficient Galaxies: Sustaining Star Formation with Stellar Mass Loss}",
    journal = {\apj},
    year = 2011,
    month = jun,
    volume = {734},
    number = {1},
    eid = {48},
    pages = {48},
    doi = {10.1088/0004-637X/734/1/48}
}

@ARTICLE{Lehnert2015,
    author = {{Lehnert}, M.~D. and {van Driel}, W. and {Le Tiran}, L. and {Di Matteo}, P. and {Haywood}, M.},
    title = "{On the cosmic evolution of the specific star formation rate}",
    journal = {\aap},
    year = 2015,
    month = may,
    volume = {577},
    eid = {A112},
    pages = {A112},
    doi = {10.1051/0004-6361/201322630}
}

@ARTICLE{Linsky1979ApJS...41...47L,
    author = {{Linsky}, J.~L. and {Worden}, S.~P. and {McClintock}, W. and {Robertson}, R.~M.},
    title = "{Stellar model chromospheres. X. High-resolution, absolute flux profiles of the Ca II H and K lines in stars of spectral types F0 - M2.}",
    journal = {\apjs},
    year = 1979,
    month = sep,
    volume = {41},
    pages = {47-74},
    doi = {10.1086/190607},
    adsurl = {https://ui.adsabs.harvard.edu/abs/1979ApJS...41...47L}
}

@ARTICLE{Maraston2005,
    author = {{Maraston}, Claudia},
    title = "{Evolutionary population synthesis: models, analysis of the ingredients and application to high-z galaxies}",
    journal = {\mnras},
    year = 2005,
    month = sep,
    volume = {362},
    number = {3},
    pages = {799-825},
    doi = {10.1111/j.1365-2966.2005.09270.x},
    archivePrefix = {arXiv},
    eprint = {astro-ph/0410207},
    primaryClass = {astro-ph},
    adsurl = {https://ui.adsabs.harvard.edu/abs/2005MNRAS.362..799M}
}

@ARTICLE{Martin2005,
    author = {{Martin}, D. Christopher and {Fanson}, James and {Schiminovich}, David and {Morrissey}, Patrick and {Friedman}, Peter G. and {Barlow}, Tom A. and {Conrow}, Tim and {Grange}, Robert and {Jelinsky}, Patrick N. and {Milliard}, Bruno and {Siegmund}, Oswald H.~W. and {Bianchi}, Luciana and {Byun}, Yong-Ik and {Donas}, Jose and {Forster}, Karl and {Heckman}, Timothy M. and {Lee}, Young-Wook and {Madore}, Barry F. and {Malina}, Roger F. and {Neff}, Susan G. and {Rich}, R. Michael and {Small}, Todd and {Surber}, Frank and {Szalay}, Alex S. and {Welsh}, Barry and {Wyder}, Ted K.},
    title = "{The Galaxy Evolution Explorer: A Space Ultraviolet Survey Mission}",
    journal = {\apjl},
    year = 2005,
    month = jan,
    volume = {619},
    number = {1},
    pages = {L1-L6},
    doi = {10.1086/426387}
}

@book{McLachlan2000,
    title = "Finite mixture models",
    author = "McLachlan, Geoffrey J. and Peel, David",
    series = "Wiley series in probability and statistics",
    publisher = "J. Wiley \& Sons",
    address = "New York",
    url = "http://opac.inria.fr/record=b1097397",
    isbn = "0-471-00626-2",
    year = 2000
}

@ARTICLE{Mints2017,
    author = {{Mints}, Alexey and {Hekker}, Saskia},
    title = "{A Unified tool to estimate Distances, Ages, and Masses (UniDAM) from spectrophotometric data}",
    journal = {\aap},
    year = 2017,
    month = aug,
    volume = {604},
    eid = {A108},
    pages = {A108},
    doi = {10.1051/0004-6361/201630090}
}

@ARTICLE{Mints2018,
    author = {{Mints}, Alexey and {Hekker}, Saskia},
    title = "{Isochrone fitting in the Gaia era}",
    journal = {\aap},
    year = 2018,
    month = oct,
    volume = {618},
    eid = {A54},
    pages = {A54},
    doi = {10.1051/0004-6361/201832739}
}

@book{Murphy2013,
    address = {Cambridge, Mass. [u.a.]},
    author = {Murphy, Kevin P.},
    description = {Machine Learning: A Probabilistic Perspective (Adaptive Computation and Machine Learning series): Kevin P. Murphy: 9780262018029: Amazon.com: Books},
    isbn = {9780262018029 0262018020},
    publisher = {MIT Press},
    refid = {904442949},
    title = {Machine learning : a probabilistic perspective},
    year = 2013
}

@ARTICLE{Nayak2024,
    author = {{Nayak}, Prasanta K. and {Ganguly}, Anindya and {Chatterjee}, Sourav},
    title = "{Hunting down white dwarf-main sequence binaries using multiwavelength observations}",
    journal = {\mnras},
    year = 2024,
    month = jan,
    volume = {527},
    number = {3},
    pages = {6100-6109},
    doi = {10.1093/mnras/stad3580},
    archivePrefix = {arXiv},
    eprint = {2212.09800},
    primaryClass = {astro-ph.SR},
    adsurl = {https://ui.adsabs.harvard.edu/abs/2024MNRAS.527.6100N}
}

@ARTICLE{Oconnell1999,
    author = {{O'Connell}, Robert W.},
    title = "{Far-Ultraviolet Radiation from Elliptical Galaxies}",
    journal = {\araa},
    year = 1999,
    month = jan,
    volume = {37},
    pages = {603-648},
    doi = {10.1146/annurev.astro.37.1.603},
    archivePrefix = {arXiv},
    eprint = {astro-ph/9906068},
    primaryClass = {astro-ph},
    adsurl = {https://ui.adsabs.harvard.edu/abs/1999ARA&A..37..603O}
}

@INPROCEEDINGS{Offner2023,
    author = {{Offner}, S.~S.~R. and {Moe}, M. and {Kratter}, K.~M. and {Sadavoy}, S.~I. and {Jensen}, E.~L.~N. and {Tobin}, J.~J.},
    title = "{The Origin and Evolution of Multiple Star Systems}",
    booktitle = {Protostars and Planets VII},
    year = 2023,
    editor = {{Inutsuka}, S. and {Aikawa}, Y. and {Muto}, T. and {Tomida}, K. and {Tamura}, M.},
    series = {Astronomical Society of the Pacific Conference Series},
    volume = {534},
    month = jul,
    pages = {275},
    doi = {10.48550/arXiv.2203.10066}
}

@ARTICLE{Parsons2016,
    author = {{Parsons}, S.~G. and {Rebassa-Mansergas}, A. and {Schreiber}, M.~R. and {G{\"a}nsicke}, B.~T. and {Zorotovic}, M. and {Ren}, J.~J.},
    title = "{The white dwarf binary pathways survey - I. A sample of FGK stars with white dwarf companions}",
    journal = {\mnras},
    year = 2016,
    month = dec,
    volume = {463},
    number = {2},
    pages = {2125-2136},
    doi = {10.1093/mnras/stw2143},
    archivePrefix = {arXiv},
    eprint = {1604.01613},
    primaryClass = {astro-ph.SR},
    adsurl = {https://ui.adsabs.harvard.edu/abs/2016MNRAS.463.2125P}
}

@ARTICLE{Peacock2018,
    author = {{Peacock}, Mark B. and {Zepf}, Stephen E. and {Maccarone}, Thomas J. and {Kundu}, Arunav and {Knigge}, Christian and {Dieball}, Andrea and {Strader}, Jay},
    title = "{Hubble Space Telescope FUV observations of M31's globular clusters suggest a spatially homogeneous helium-enriched subpopulation}",
    journal = {\mnras},
    year = 2018,
    month = dec,
    volume = {481},
    number = {3},
    pages = {3313-3324},
    doi = {10.1093/mnras/sty2461}
}

@ARTICLE{Piau2002,
    author = {{Piau}, L. and {Turck-Chi{\`e}ze}, S.},
    title = "{Lithium Depletion in Pre-Main-Sequence Solar-like Stars}",
    journal = {\apj},
    year = 2002,
    month = feb,
    volume = {566},
    number = {1},
    pages = {419-434},
    doi = {10.1086/324277},
    archivePrefix = {arXiv},
    eprint = {astro-ph/0111223},
    primaryClass = {astro-ph},
    adsurl = {https://ui.adsabs.harvard.edu/abs/2002ApJ...566..419P}
}

@ARTICLE{Psaradaki2024,
    author = {{Psaradaki}, I. and {Corrales}, L. and {Werk}, J. and {Jensen}, A.~G. and {Costantini}, E. and {Mehdipour}, M. and {Cilley}, R. and {Schulz}, N. and {Kaastra}, J. and {Garc{\'\i}a}, J.~A. and {Valencic}, L. and {Kallman}, T. and {Paerels}, F.},
    title = "{Elemental Abundances in the Diffuse Interstellar Medium from Joint Far-ultraviolet and X-Ray Spectroscopy: Iron, Oxygen, Carbon, and Sulfur}",
    journal = {\aj},
    year = 2024,
    month = may,
    volume = {167},
    number = {5},
    eid = {217},
    pages = {217},
    doi = {10.3847/1538-3881/ad306b}
}

@ARTICLE{Queiroz2018,
    author = {{Queiroz}, A.~B.~A. and {Anders}, F. and {Santiago}, B.~X. and {Chiappini}, C. and {Steinmetz}, M. and {Dal Ponte}, M. and {Stassun}, K.~G. and {da Costa}, L.~N. and {Maia}, M.~A.~G. and {Crestani}, J. and {Beers}, T.~C. and {Fern{\'a}ndez-Trincado}, J.~G. and {Garc{\'\i}a-Hern{\'a}ndez}, D.~A. and {Roman-Lopes}, A. and {Zamora}, O.},
    title = "{StarHorse: a Bayesian tool for determining stellar masses, ages, distances, and extinctions for field stars}",
    journal = {\mnras},
    year = 2018,
    month = may,
    volume = {476},
    number = {2},
    pages = {2556-2583},
    doi = {10.1093/mnras/sty330}
}

@ARTICLE{Queiroz2023,
    author = {{Queiroz}, A.~B.~A. and {Anders}, F. and {Chiappini}, C. and {Khalatyan}, A. and {Santiago}, B.~X. and {Nepal}, S. and {Steinmetz}, M. and {Gallart}, C. and {Valentini}, M. and {Dal Ponte}, M. and {Barbuy}, B. and {P{\'e}rez-Villegas}, A. and {Masseron}, T. and {Fern{\'a}ndez-Trincado}, J.~G. and {Khoperskov}, S. and {Minchev}, I. and {Fern{\'a}ndez-Alvar}, E. and {Lane}, R.~R. and {Nitschelm}, C.},
    title = "{StarHorse results for spectroscopic surveys and Gaia DR3: Chrono-chemical populations in the solar vicinity, the genuine thick disk, and young alpha-rich stars}",
    journal = {\aap},
    year = 2023,
    month = may,
    volume = {673},
    eid = {A155},
    pages = {A155},
    doi = {10.1051/0004-6361/202245399}
}

@ARTICLE{Randich2022,
    author = {{Randich}, S. and {Gilmore}, G. and {Magrini}, L. and {Sacco}, G.~G. and {Jackson}, R.~J. and {Jeffries}, R.~D. and {Worley}, C.~C. and others},
    title = "{The Gaia-ESO Public Spectroscopic Survey: Implementation, data products, open cluster survey, science, and legacy}",
    journal = {\aap},
    year = 2022,
    month = oct,
    volume = {666},
    eid = {A121},
    pages = {A121},
    doi = {10.1051/0004-6361/202243141}
}

@ARTICLE{Raso2017,
    author = {{Raso}, S. and {Ferraro}, F.~R. and {Dalessandro}, E. and {Lanzoni}, B. and {Nardiello}, D. and {Bellini}, A. and {Vesperini}, E.},
    title = "{The {\textquotedblleft}UV-route{\textquotedblright} to Search for Blue Straggler Stars in Globular Clusters: First Results from the HST UV Legacy Survey}",
    journal = {\apj},
    year = 2017,
    month = apr,
    volume = {839},
    number = {1},
    eid = {64},
    pages = {64},
    doi = {10.3847/1538-4357/aa6891}
}

@ARTICLE{Reggiani2025,
    author = {{Reggiani}, Elisabetta and {Cadelano}, Mario and {Lanzoni}, Barbara and {Ferraro}, Francesco R. and {Salaris}, Maurizio and {Mucciarelli}, Alessio},
    title = "{Detection of a white dwarf orbiting a carbon-oxygen-depleted blue straggler in 47 Tucanae}",
    journal = {\aap},
    year = 2025,
    month = oct,
    volume = {702},
    eid = {A185},
    pages = {A185},
    doi = {10.1051/0004-6361/202556218},
    archivePrefix = {arXiv},
    eprint = {2508.21118},
    primaryClass = {astro-ph.SR},
    adsurl = {https://ui.adsabs.harvard.edu/abs/2025A&A...702A.185R}
}

@ARTICLE{RixBovy2013,
    author = {{Rix}, Hans-Walter and {Bovy}, Jo},
    title = "{The Milky Way's stellar disk. Mapping and modeling the Galactic disk}",
    journal = {\aapr},
    year = 2013,
    month = may,
    volume = {21},
    eid = {61},
    pages = {61},
    doi = {10.1007/s00159-013-0061-8}
}

@software{RohatgiWebPlotDigitizer,
    author  = {Rohatgi, Ankit},
    title   = {WebPlotDigitizer},
    year    = {2024},
    version = {5.2},
    url     = {https://automeris.io},
}

@ARTICLE{Rosenfield2012,
    author = {{Rosenfield}, Philip and {Johnson}, L. Clifton and {Girardi}, L{\'e}o and {Dalcanton}, Julianne J. and {Bressan}, Alessandro and {Lang}, Dustin and {Williams}, Benjamin F. and {Guhathakurta}, Puragra and {Howley}, Kirsten M. and {Lauer}, Tod R. and {Bell}, Eric F. and {Bianchi}, Luciana and {Caldwell}, Nelson and {Dolphin}, Andrew and {Dorman}, Claire E. and {Gilbert}, Karoline M. and {Kalirai}, Jason and {Larsen}, S{\o}ren S. and {Olsen}, Knut A.~G. and {Rix}, Hans-Walter and {Seth}, Anil C. and {Skillman}, Evan D. and {Weisz}, Daniel R.},
    title = "{The Panchromatic Hubble Andromeda Treasury. I. Bright UV Stars in the Bulge of M31}",
    journal = {\apj},
    year = 2012,
    month = aug,
    volume = {755},
    number = {2},
    eid = {131},
    pages = {131},
    doi = {10.1088/0004-637X/755/2/131}
}

@ARTICLE{Salpeter1955,
    author = {{Salpeter}, Edwin E.},
    title = "{The Luminosity Function and Stellar Evolution.}",
    journal = {\apj},
    year = 1955,
    month = jan,
    volume = {121},
    pages = {161},
    doi = {10.1086/145971},
    adsurl = {https://ui.adsabs.harvard.edu/abs/1955ApJ...121..161S}
}

@ARTICLE{SalvadorRusinol2020,
    author = {{Salvador-Rusi{\~n}ol}, N{\'u}ria and {Vazdekis}, Alexandre and {La Barbera}, Francesco and {Beasley}, Michael A. and {Ferreras}, Ignacio and {Negri}, Andrea and {Dalla Vecchia}, Claudio},
    title = "{Sub one per cent mass fractions of young stars in red massive galaxies}",
    journal = {Nature Astronomy},
    year = 2020,
    month = jan,
    volume = {4},
    pages = {252-259},
    doi = {10.1038/s41550-019-0955-0},
    archivePrefix = {arXiv},
    eprint = {1912.06700},
    primaryClass = {astro-ph.GA},
    adsurl = {https://ui.adsabs.harvard.edu/abs/2020NatAs...4..252S}
}

@ARTICLE{Santos2013,
    author = {{Santos}, N.~C. and {Sousa}, S.~G. and {Mortier}, A. and {Neves}, V. and {Adibekyan}, V. and {Tsantaki}, M. and {Delgado Mena}, E. and {Bonfils}, X. and {Israelian}, G. and {Mayor}, M. and {Udry}, S.},
    title = "{SWEET-Cat: A catalogue of parameters for Stars With ExoplanETs. I. New atmospheric parameters and masses for 48 stars with planets}",
    journal = {\aap},
    year = 2013,
    month = aug,
    volume = {556},
    eid = {A150},
    pages = {A150},
    doi = {10.1051/0004-6361/201321286},
    archivePrefix = {arXiv},
    eprint = {1307.0354},
    primaryClass = {astro-ph.SR},
    adsurl = {https://ui.adsabs.harvard.edu/abs/2013A&A...556A.150S}
}

@article{Schwarz1978,
    title = {Estimating the Dimension of a Model},
    volume = {6},
    ISSN = {0090-5364},
    url = {http://dx.doi.org/10.1214/aos/1176344136},
    DOI = {10.1214/aos/1176344136},
    number = {2},
    journal = {The Annals of Statistics},
    publisher = {Institute of Mathematical Statistics},
    author = {Schwarz,  Gideon},
    year = {1978},
    month = Mar 
}

@ARTICLE{Shkolnik2011,
    author = {{Shkolnik}, Evgenya L. and {Liu}, Michael C. and {Reid}, I. Neill and {Dupuy}, Trent and {Weinberger}, Alycia J.},
    title = "{Searching for Young M Dwarfs with GALEX}",
    journal = {\apj},
    year = 2011,
    month = jan,
    volume = {727},
    number = {1},
    eid = {6},
    pages = {6},
    doi = {10.1088/0004-637X/727/1/6},
    archivePrefix = {arXiv},
    eprint = {1011.2708},
    primaryClass = {astro-ph.SR},
    adsurl = {https://ui.adsabs.harvard.edu/abs/2011ApJ...727....6S}
}

@ARTICLE{Shkolnik2014,
    author = {{Shkolnik}, Evgenya L. and {Barman}, Travis S.},
    title = "{HAZMAT. I. The Evolution of Far-UV and Near-UV Emission from Early M Stars}",
    journal = {\aj},
    year = 2014,
    month = oct,
    volume = {148},
    number = {4},
    eid = {64},
    pages = {64},
    doi = {10.1088/0004-6256/148/4/64}
}

@ARTICLE{Sindhu2018,
    author = {{Sindhu}, N. and {Subramaniam}, Annapurni and {Radha}, C. Anu},
    title = "{Ultraviolet stellar population of the old open cluster M67 (NGC 2682)}",
    journal = {\mnras},
    year = 2018,
    month = nov,
    volume = {481},
    number = {1},
    pages = {226-243},
    doi = {10.1093/mnras/sty2283},
    archivePrefix = {arXiv},
    eprint = {1808.06814},
    primaryClass = {astro-ph.SR},
    adsurl = {https://ui.adsabs.harvard.edu/abs/2018MNRAS.481..226S}
}

@ARTICLE{Skrutskie2006,
    author = {{Skrutskie}, M.~F. and {Cutri}, R.~M. and {Stiening}, R. and {Weinberg}, M.~D. and {Schneider}, S. and {Carpenter}, J.~M. and {Beichman}, C. and {Capps}, R. and {Chester}, T. and {Elias}, J. and {Huchra}, J. and {Liebert}, J. and {Lonsdale}, C. and {Monet}, D.~G. and {Price}, S. and {Seitzer}, P. and {Jarrett}, T. and {Kirkpatrick}, J.~D. and {Gizis}, J.~E. and {Howard}, E. and {Evans}, T. and {Fowler}, J. and {Fullmer}, L. and {Hurt}, R. and {Light}, R. and {Kopan}, E.~L. and {Marsh}, K.~A. and {McCallon}, H.~L. and {Tam}, R. and {Van Dyk}, S. and {Wheelock}, S.},
    title = "{The Two Micron All Sky Survey (2MASS)}",
    journal = {\aj},
    year = 2006,
    month = feb,
    volume = {131},
    number = {2},
    pages = {1163-1183},
    doi = {10.1086/498708},
    adsurl = {https://ui.adsabs.harvard.edu/abs/2006AJ....131.1163S}
}

@ARTICLE{Smith2010,
    author = {{Smith}, Graeme H. and {Redenbaugh}, Anja K.},
    title = "{A Dependence of GALEX FUV Magnitudes of F, G, and K Dwarfs Upon Stellar Activity}",
    journal = {\pasp},
    year = 2010,
    month = nov,
    volume = {122},
    number = {897},
    pages = {1303},
    doi = {10.1086/657051},
    adsurl = {https://ui.adsabs.harvard.edu/abs/2010PASP..122.1303S}
}

@ARTICLE{Smith2014,
    author = {{Smith}, Myron A. and {Bianchi}, Luciana and {Shiao}, Bernard},
    title = "{Interesting Features in the Combined GALEX and Sloan Color Diagrams of Solar-like Galactic Populations}",
    journal = {\aj},
    year = 2014,
    month = jun,
    volume = {147},
    number = {6},
    eid = {159},
    pages = {159},
    doi = {10.1088/0004-6256/147/6/159}
}

@ARTICLE{Souza2024MNRAS.532..563S,
    author = {{Souza dos Santos}, P.~V. and {Porto de Mello}, G.~F. and {Costa-Bhering}, E. and {Lorenzo-Oliveira}, D. and {Almeida-Fernandes}, F. and {Dutra-Ferreira}, L. and {Ribas}, I.},
    title = "{Fine structure of the age-chromospheric activity relation in solar-type stars: II. H{\ensuremath{\alpha}} line}",
    journal = {\mnras},
    year = 2024,
    month = jul,
    volume = {532},
    number = {1},
    pages = {563-576},
    doi = {10.1093/mnras/stae1532},
    archivePrefix = {arXiv},
    eprint = {2406.12519},
    primaryClass = {astro-ph.SR},
    adsurl = {https://ui.adsabs.harvard.edu/abs/2024MNRAS.532..563S}
}

@article{Spearman1904,
    ISSN = {00029556},
    URL = {http://www.jstor.org/stable/1412159},
    author = {C. Spearman},
    journal = {The American Journal of Psychology},
    number = {1},
    pages = {72--101},
    publisher = {University of Illinois Press},
    title = {The Proof and Measurement of Association between Two Things},
    urldate = {2023-11-09},
    volume = {15},
    year = {1904}
}

@ARTICLE{Stonkunte2016,
    author = {{Stonkut{\.{e}}}, E. and {Koposov}, S.~E. and {Howes}, L.~M. and {Feltzing}, S. and {Worley}, C.~C. and {Gilmore}, G. and {Ruchti}, G.~R. and {Kordopatis}, G. and {Randich}, S. and {Zwitter}, T. and {Bensby}, T. and {Bragaglia}, A. and {Smiljanic}, R. and {Costado}, M.~T. and {Tautvai{\v{s}}ien{\.{e}}}, G. and {Casey}, A.~R. and {Korn}, A.~J. and {Lanzafame}, A.~C. and {Pancino}, E. and {Franciosini}, E. and {Hourihane}, A. and {Jofr{\'e}}, P. and {Lardo}, C. and {Lewis}, J. and {Magrini}, L. and {Monaco}, L. and {Morbidelli}, L. and {Sacco}, G.~G. and {Sbordone}, L.},
    title = "{The Gaia-ESO Survey: the selection function of the Milky Way field stars}",
    journal = {\mnras},
    year = 2016,
    month = jul,
    volume = {460},
    number = {1},
    pages = {1131-1146},
    doi = {10.1093/mnras/stw1011}
}

@ARTICLE{Vincenzo2016,
    author = {{Vincenzo}, F. and {Matteucci}, F. and {Belfiore}, F. and {Maiolino}, R.},
    title = "{Modern yields per stellar generation: the effect of the IMF}",
    journal = {\mnras},
    year = 2016,
    month = feb,
    volume = {455},
    number = {4},
    pages = {4183-4190},
    doi = {10.1093/mnras/stv2598},
    archivePrefix = {arXiv},
    eprint = {1503.08300},
    primaryClass = {astro-ph.GA},
    adsurl = {https://ui.adsabs.harvard.edu/abs/2016MNRAS.455.4183V}
}

@ARTICLE{Xin2007,
    author = {{Xin}, Y. and {Deng}, L. and {Han}, Z.~W.},
    title = "{Blue Straggler Stars in Galactic Open Clusters and the Simple Stellar Population Model}",
    journal = {\apj},
    year = 2007,
    month = may,
    volume = {660},
    number = {1},
    pages = {319-329},
    doi = {10.1086/512964}
}

@ARTICLE{Yi2011,
    author = {{Yi}, Sukyoung K. and {Lee}, Jihye and {Sheen}, Yun-Kyeong and {Jeong}, Hyunjin and {Suh}, Hyewon and {Oh}, Kyuseok},
    title = "{The Ultraviolet Upturn in Elliptical Galaxies and Environmental Effects}",
    journal = {\apjs},
    year = 2011,
    month = aug,
    volume = {195},
    number = {2},
    eid = {22},
    pages = {22},
    doi = {10.1088/0067-0049/195/2/22}
}

@ARTICLE{Wade1998,
    author = {{Wade}, Richard A. and {Hubeny}, Ivan},
    title = "{Detailed Mid- and Far-Ultraviolet Model Spectra for Accretion Disks in Cataclysmic Binaries}",
    journal = {\apj},
    year = 1998,
    month = dec,
    volume = {509},
    number = {1},
    pages = {350-361},
    doi = {10.1086/306496}
}

@ARTICLE{Walcher2011,
    author = {{Walcher}, Jakob and {Groves}, Brent and {Budav{\'a}ri}, Tam{\'a}s and {Dale}, Daniel},
    title = "{Fitting the integrated spectral energy distributions of galaxies}",
    journal = {\apss},
    year = 2011,
    month = jan,
    volume = {331},
    number = {1},
    pages = {1-51},
    doi = {10.1007/s10509-010-0458-z}
}

@ARTICLE{Weiss2006,
    author = {{Weiss}, A. and {Salaris}, M. and {Ferguson}, J.~W. and {Alexander}, D.~R.},
    title = "{alpha-element enhanced opacity tables and low-mass metal-rich stellar models}",
    journal = {arXiv e-prints},
    year = 2006,
    month = may,
    eid = {astro-ph/0605666},
    pages = {astro-ph/0605666},
    doi = {10.48550/arXiv.astro-ph/0605666},
    archivePrefix = {arXiv},
    eprint = {astro-ph/0605666},
    primaryClass = {astro-ph},
    adsurl = {https://ui.adsabs.harvard.edu/abs/2006astro.ph..5666W}
}

@article{Werle2019,
    author = {Werle, A. and {Cid Fernandes}, R. and Asari, N. Vale and Bruzual, G. and Charlot, S. and {Gonzalez Delgado}, R. and Herpich, F. R.},
    doi = {10.1093/mnras/sty3264},
    eprint = {1811.11255},
    issn = {13652966},
    journal = {\mnras},
    number = {2},
    pages = {2382--2397},
    publisher = {Oxford University Press},
    title = {{Simultaneous analysis of SDSS spectra and GALEX photometry with STARLIGHT: Method and early results}},
    volume = {483},
    year = {2019}
}

@article{Wien1893,
    title = {Die obere Grenze der Wellenl\"{a}ngen,  welche in der W\"{a}rmestrahlung fester K\"{o}rper vorkommen k\"{o}nnen; Folgerungen aus dem zweiten Hauptsatz der W\"{a}rmetheorie},
    volume = {285},
    ISSN = {1521-3889},
    url = {http://dx.doi.org/10.1002/andp.18932850803},
    DOI = {10.1002/andp.18932850803},
    number = {8},
    journal = {Annalen der Physik},
    publisher = {Wiley},
    author = {Wien,  Willy},
    year = {1893},
    month = jan,
    pages = {633–641}
}

@ARTICLE{York2000,
    author = {{York}, Donald G. and {Adelman}, J. and {Anderson}, John E., Jr. and {Anderson}, Scott F. and {Annis}, James and {Bahcall}, Neta A. and {Bakken}, J.~A. and {Barkhouser}, Robert and {Bastian}, Steven and {Berman}, Eileen and {Boroski}, William N. and {Bracker}, Steve and {Briegel}, Charlie and {Briggs}, John W. and {Brinkmann}, J. and {Brunner}, Robert and {Burles}, Scott and {Carey}, Larry and {Carr}, Michael A. and {Castander}, Francisco J. and {Chen}, Bing and {Colestock}, Patrick L. and {Connolly}, A.~J. and {Crocker}, J.~H. and {Csabai}, Istv{\'a}n and {Czarapata}, Paul C. and {Davis}, John Eric and {Doi}, Mamoru and {Dombeck}, Tom and {Eisenstein}, Daniel and {Ellman}, Nancy and {Elms}, Brian R. and {Evans}, Michael L. and {Fan}, Xiaohui and {Federwitz}, Glenn R. and {Fiscelli}, Larry and {Friedman}, Scott and {Frieman}, Joshua A. and {Fukugita}, Masataka and {Gillespie}, Bruce and {Gunn}, James E. and {Gurbani}, Vijay K. and {de Haas}, Ernst and {Haldeman}, Merle and {Harris}, Frederick H. and {Hayes}, J. and {Heckman}, Timothy M. and {Hennessy}, G.~S. and {Hindsley}, Robert B. and {Holm}, Scott and {Holmgren}, Donald J. and {Huang}, Chi-hao and {Hull}, Charles and {Husby}, Don and {Ichikawa}, Shin-Ichi and {Ichikawa}, Takashi and {Ivezi{\'c}}, {\v{Z}}eljko and {Kent}, Stephen and {Kim}, Rita S.~J. and {Kinney}, E. and {Klaene}, Mark and {Kleinman}, A.~N. and {Kleinman}, S. and {Knapp}, G.~R. and {Korienek}, John and {Kron}, Richard G. and {Kunszt}, Peter Z. and {Lamb}, D.~Q. and {Lee}, B. and {Leger}, R. French and {Limmongkol}, Siriluk and {Lindenmeyer}, Carl and {Long}, Daniel C. and {Loomis}, Craig and {Loveday}, Jon and {Lucinio}, Rich and {Lupton}, Robert H. and {MacKinnon}, Bryan and {Mannery}, Edward J. and {Mantsch}, P.~M. and {Margon}, Bruce and {McGehee}, Peregrine and {McKay}, Timothy A. and {Meiksin}, Avery and {Merelli}, Aronne and {Monet}, David G. and {Munn}, Jeffrey A. and {Narayanan}, Vijay K. and {Nash}, Thomas and {Neilsen}, Eric and {Neswold}, Rich and {Newberg}, Heidi Jo and {Nichol}, R.~C. and {Nicinski}, Tom and {Nonino}, Mario and {Okada}, Norio and {Okamura}, Sadanori and {Ostriker}, Jeremiah P. and {Owen}, Russell and {Pauls}, A. George and {Peoples}, John and {Peterson}, R.~L. and {Petravick}, Donald and {Pier}, Jeffrey R. and {Pope}, Adrian and {Pordes}, Ruth and {Prosapio}, Angela and {Rechenmacher}, Ron and {Quinn}, Thomas R. and {Richards}, Gordon T. and {Richmond}, Michael W. and {Rivetta}, Claudio H. and {Rockosi}, Constance M. and {Ruthmansdorfer}, Kurt and {Sandford}, Dale and {Schlegel}, David J. and {Schneider}, Donald P. and {Sekiguchi}, Maki and {Sergey}, Gary and {Shimasaku}, Kazuhiro and {Siegmund}, Walter A. and {Smee}, Stephen and {Smith}, J. Allyn and {Snedden}, S. and {Stone}, R. and {Stoughton}, Chris and {Strauss}, Michael A. and {Stubbs}, Christopher and {SubbaRao}, Mark and {Szalay}, Alexander S. and {Szapudi}, Istvan and {Szokoly}, Gyula P. and {Thakar}, Anirudda R. and {Tremonti}, Christy and {Tucker}, Douglas L. and {Uomoto}, Alan and {Vanden Berk}, Dan and {Vogeley}, Michael S. and {Waddell}, Patrick and {Wang}, Shu-i. and {Watanabe}, Masaru and {Weinberg}, David H. and {Yanny}, Brian and {Yasuda}, Naoki and {SDSS Collaboration}},
    title = "{The Sloan Digital Sky Survey: Technical Summary}",
    journal = {\aj},
    year = 2000,
    month = sep,
    volume = {120},
    number = {3},
    pages = {1579-1587},
    doi = {10.1086/301513}
}

\begin{appendix}
\nolinenumbers
\onecolumn

\section{Supplementary quantities for the stellar and extragalactic analyses}
\label{app:supplementary}

This appendix provides the numerical quantities supporting the comparison of the \tgex\ stellar groups and the empirical UV-luminosity scaling presented in the main text. Table~\ref{tab:gmms_params} summarises the stellar-parameter distributions associated with the GMM classification, while Tables~\ref{tab:selected_uvupturn_galaxies} and \ref{tab:tgex_uv_templates} list the galaxy and stellar inputs adopted in the extragalactic feasibility calculation.

\subsection{Stellar properties of the GMM groups}
\label{app:gmm_properties}

Table~\ref{tab:gmms_params} reports the 2.3rd, 16th, 50th, 84th, and 97.7th percentiles of the \teff, age, \feh, and \mgfe\ distributions shown in Fig.~\ref{fig:params_distribution}. These percentiles provide non-parametric summaries of the central values and dispersion within each GMM-defined UV-colour group. Group~1 occupies a narrow, cool \teff\ range and has the highest median \mgfe, whereas Group~2 contains the hottest stars and spans the broadest metallicity range. Group~3 has intermediate temperatures and the highest median \feh. The age distributions are broad, particularly for Groups~1 and 3, and should be interpreted in view of the isochrone degeneracies and small group sizes discussed in the main text. The tabulated values describe the observed \tgex\ sample and are not intended as estimates of the intrinsic distributions of UV-normal or UV-abnormal Galactic stars.

\newcommand{\ta}[1]{\multicolumn{1}{c}{\hspace{1.3em}#1}}
\begin{table}[ht!]
    \centering
    \caption{Descriptive statistics of the stellar parameters depicted in Fig. \ref{fig:params_distribution}.}
    \label{tab:gmms_params}
    \renewcommand{\arraystretch}{1.15}
    \begin{tabular}{c l S[table-format=4.2] S[table-format=4.2]  S[table-format=4.2] S[table-format=4.2]  S[table-format=4.2]}
        \toprule
        GMM group & Parameter & \ta{2.3\%} & \ta{16\%} & \ta{50\%} & \ta{84\%} & \ta{97.7\%} \\
                  & & \ta{$-2\sigma$} & \ta{$-1\sigma$} & \ta{median} & \ta{$+1\sigma$} & \ta{$+2\sigma$} \\
        \midrule

        \multirow{4}{*}{1 (red)}
        & \teff\ (K)    &  \ta{4795}  &  \ta{4826}  &  \ta{4889} &  \ta{4937} &   \ta{4978} \\
        & Age (Gyr)     &  0.03  &  0.03  &  5.89 & 11.16 &  12.14 \\
        & \feh\         & -0.63  & -0.58  & -0.52 & -0.23 &  -0.20  \\
        & \mgfe\        &  0.03  &  0.12  &  0.26 &  0.40 &   0.40  \\
        \midrule

        \multirow{4}{*}{2 (blue)}
        & \teff\ (K)    &  \ta{6203} &  \ta{6267} &  \ta{6546} &  \ta{6782} & \ta{6833} \\
        & Age (Gyr)     &  1.33 &  2.39 &  4.27 &  7.71 & 8.32 \\
        & \feh\         & -1.34 & -1.23 & -0.47 & -0.19 & 0.03 \\
        & \mgfe\        & -0.04 &  0.04 &  0.09 &  0.27 & 0.54 \\
        \midrule

        \multirow{4}{*}{3 (yellow)}
        & \teff\ (K)    &  \ta{4829} &  \ta{5031} &  \ta{5752} &  \ta{5943} &  \ta{6618} \\
        & Age (Gyr)     &  0.01 &  0.02 &  5.89 & 12.30 & 13.49 \\
        & \feh\         & -0.72 & -0.50 & -0.09 &  0.13 &  0.23 \\
        & \mgfe\        & -0.14 & -0.02 &  0.07 &  0.34 &  0.43 \\
        \bottomrule
    \end{tabular}
\end{table}

\subsection{Inputs to the extragalactic UV-luminosity scaling}
\label{app:scaling_inputs}

Table~\ref{tab:selected_uvupturn_galaxies} lists the three representative quiescent galaxies used in the feasibility calculation. The galaxies were selected from the low-, intermediate-, and high-UV-brightness terciles of the parent UV-upturn sample and are used to probe the scaling across systems with different observed UV outputs. The table gives their redshifts, absolute GALEX magnitudes, present-day catalogue stellar masses, inferred formed stellar masses, adopted population ages, and fiducial numbers of surviving stars in the restricted FGK mass interval. The selected systems have similar representative ages but differ substantially in stellar mass and UV luminosity, producing surviving FGK populations of approximately $(1.1$--$3.3)\times10^{10}$ stars.

Table~\ref{tab:tgex_uv_templates} gives the three empirical Group~1 templates used to assign per-star FUV and NUV luminosities to the hypothetical UV-excess FGK component. The red and blue templates correspond to the Group~1 stars with the reddest and bluest FUV--NUV colours, respectively, while the median template is constructed independently from the median Group~1 FUV and NUV absolute magnitudes. The large luminosity range between these templates drives the strongly non-linear results shown in Fig.~\ref{fig:nuv_fuv_contributions}: the median template produces a limited integrated contribution, whereas the bluest template represents an extreme UV-luminosity case in which a comparatively small FGK fraction can match a substantial part of the observed galaxy UV output. The templates are empirical luminosity prescriptions and are not interpreted as complete stellar SEDs or as evidence that the corresponding UV behaviour is intrinsic to the optically dominant FGK star.

\begin{table}[!ht]
    \centering
    \caption{Representative UV-upturn galaxies used in the empirical scaling calculation. Present-day stellar masses are taken from the GAMA catalogue; formed stellar masses assume $R=0.4$ via Eq. 2.
    The final column gives the number of surviving stars in the adopted FGK mass window, computed using Eq. 5 with the catalogue age and fiducial $\eta=2.5$.
    }
    \label{tab:selected_uvupturn_galaxies}
    \begin{tabular}{lcccccccc}
        \toprule
        CATAID & UV-score tercile & $z$ & $M_{\rm FUV}$ & $M_{\rm NUV}$ & $M_{\star}^{\rm present}$ & $M_{\star}^{\rm formed}$ & $\tau$ & $N_{\rm FGK}^{\rm surv}$ \\
        & & & mag & mag & $10^{10}\,M_\odot$ & $10^{10}\,M_\odot$ & Gyr & $10^{10}$ \\
        \midrule
        423376 & low  & 0.195 & -15.88 & -15.82 & 6.62 & 11.0 & 7.01 & 2.95 \\
        31326  & mid  & 0.190 & -15.55 & -14.89 & 7.41 & 12.4 & 7.43 & 3.31 \\
        30793  & high & 0.153 & -14.78 & -14.61 & 2.49 & 4.15 & 7.12 & 1.11 \\
        \bottomrule
    \end{tabular}
\end{table}

\begin{table}[!ht]
    \centering
    \caption{Empirical \tgex\ UV templates used in the extragalactic scaling calculation. The red and blue templates correspond to the individual \tgex\ Group~1 stars with the reddest and bluest FUV--NUV colours, respectively. The median template is constructed from the Group~1 median FUV and NUV absolute magnitudes. Luminosities are monochromatic AB luminosities.}
    \label{tab:tgex_uv_templates}
    \begin{tabular}{lcccccc}
        \toprule
        Template & Definition & $M_{\rm FUV}^{\star}$ & $M_{\rm NUV}^{\star}$ & FUV--NUV & $L_{\nu,{\rm FUV}}^{\star}$ & $L_{\nu,{\rm NUV}}^{\star}$ \\
        & & mag & mag & mag & erg s$^{-1}$ Hz$^{-1}$ & erg s$^{-1}$ Hz$^{-1}$ \\
        \midrule
        Red    & reddest star & 14.45 & 13.33 &  1.12 & $7.20\times10^{14}$ & $2.01\times10^{15}$ \\
        Median & group median & 11.92 & 12.31 & -0.39 & $7.42\times10^{15}$ & $5.16\times10^{15}$ \\
        Blue   & bluest star  &  7.79 &  8.34 & -0.55 & $3.31\times10^{17}$ & $2.01\times10^{17}$ \\
        \bottomrule
    \end{tabular}
\end{table}




\end{appendix}

\end{document}